\documentclass[amsmath,amssymb]{revtex4-1}
\usepackage{style}

\usepackage{graphicx}
\usepackage{dcolumn}
\usepackage{bm}
\usepackage{natbib}

\input epsf
\usepackage{psfrag}
\usepackage{graphicx}

\newcommand\re{{\rm Re}\,}
\newcommand\We{{\rm We}\,}

\newcommand\Ma{{\rm Ma}\,}

\newcommand\Fr{{\rm Fr}\,}
\newcommand\St{{\rm St}}

\begin{document}

\title[Anatomy of the splash]{Droplet impact on a thin liquid film: anatomy of the splash}

\author{Christophe Josserand} 
\affiliation{Sorbonne Universit\'es, UPMC Univ Paris 06, UMR 7190,
Institut Jean Le Rond d'Alembert, F-75005, Paris, France\\
CNRS, UMR 7190, Institut Jean Le Rond d'Alembert, F-75005, Paris, France}
\author{Pascal Ray}
\affiliation{Sorbonne Universit\'es, UPMC Univ Paris 06, UMR 7190,
Institut Jean Le Rond d'Alembert, F-75005, Paris, France\\
CNRS, UMR 7190, Institut Jean Le Rond d'Alembert, F-75005, Paris, France}
\author{St\'ephane Zaleski}
\affiliation{Sorbonne Universit\'es, UPMC Univ Paris 06, UMR 7190,
Institut Jean Le Rond d'Alembert, F-75005, Paris, France\\
CNRS, UMR 7190, Institut Jean Le Rond d'Alembert, F-75005, Paris, France}

\begin{abstract}
We investigate the dynamics of drop impact on a thin liquid film at short times in order to identify the
mechanisms of splashing formation. Using numerical simulations and scaling analysis, we show that 
the splashing formation depends both on the inertial dynamics of the liquid and the cushioning of the 
gas. Two asymptotic regimes are identified, characterized by a new dimensionless number $J$: 
when the gas cushioning is weak, the jet is formed after a sequence of bubbles are entrapped and the 
jet speed is mostly selected by the Reynolds number of the impact. On the other hand, when the air 
cushioning is important, the lubrication of the gas beneath the drop and the liquid film controls the 
dynamics, leading to a single bubble entrapment and a weaker jet velocity.
\end{abstract}

\maketitle

\section{Introduction}
\begin{figure}
\centering
a)\includegraphics[width=5. cm,height=2.5cm]{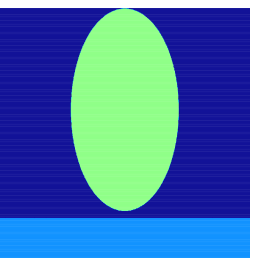}b)\includegraphics[width=5. cm,height=2.5cm]{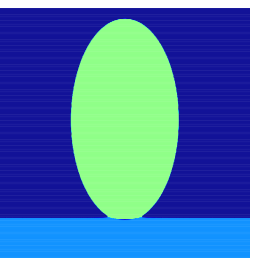}\\
c)\includegraphics[width=5. cm,height=2.5cm]{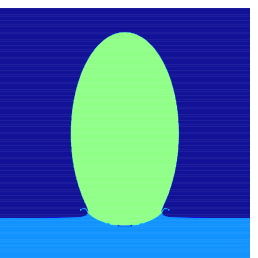}d)\includegraphics[width=5. cm,height=2.5cm]{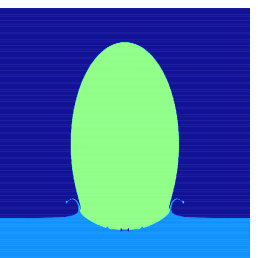}\\
e)\includegraphics[width=5. cm,height=2.5cm]{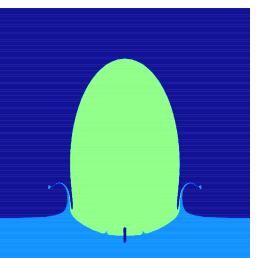}f)\includegraphics[width=5. cm,height=2.5cm]{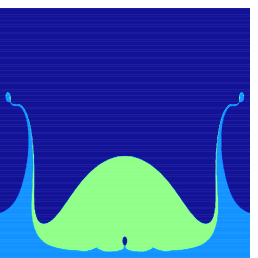}\\
\caption{Snapshots of a droplet impacting on a thin liquid film
for parameters in Tables~\ref{allparams} and \ref{tabledimsnap}, {\it i.e.} $\We=500$ and $\re=2000$.  
For the sake of visualisation, the liquid of the droplet and of the layer have been colored in green and blue respectively, although it is the same liquid. The gas phase is colored in dark blue. The snapshots 
correspond to the dimensionless time $U_0t/D=-0.033$, $0.0167$,$0.083$, $0.133$, $0.217$ and $0.767$ respectively.}
\label{snap}
\end{figure}

Droplet collision and impact are iconic multiphase flow problems:
rain, atomization of liquid jets, ink-jet printing or stalagmite
growth all involve impact in one manner or the
other~\citep{Rein93,Yarin06,JTAR16}. The droplet may impact on a dry surface,
a thin liquid film or a deep liquid bath. In all cases, 
impact may lead to the 
spreading of the droplet or to a splash
where a myriad of smaller droplets are ejected far away from the zone of
impact~\citep{Rio01}. Control of the outcome
of impact is crucial for applications. Spreading is desirable
for coating or ink-jet printing for instance while splashing 
may improve the efficiency of evaporation and mixing in combustion chambers.
Two distinct types of splash, prompt and ``ordinary'' are now
distinguished. The prompt splash is defined as a very early ejection
of liquid at a time $t \ll D/U_0$ where $D$ is the droplet diameter and $U_0$ its velocity. 
The second, ordinary type  occurs at larger times,
through the formation of a vertical corolla
ending in a circular rim that destabilizes into fingers and
droplets~\citep{DeBr08}.
In the prompt splash, a very thin liquid jet first forms, called the ejecta sheet~\citep{tho02}. 
This sheet is ejected at high velocity, initially almost horizontally and is expected to 
disintegrate eventually in very small, fast droplets. 
Very often, these two mechanisms happen in a sequence, 
the ejecta sheet being a precursor of the corolla, 
as illustrated by the numerical simulations of droplet impacts on
a thin liquid film shown in Figure \ref{snap}.  

The dynamics of droplet impact is complex, involving 
singular surface deformation and pressure values in the inviscid limit
and several instabilities of surface evolution, 
so that an overall understanding of the whole process is
still lacking.  In particular, the splash depends on many physical
parameters, the most important being the impact velocity. Obviously,
high velocities promote the splash while at low velocities the droplet gently spreads. 
This behavior is mostly characterized by
the Reynolds and the Weber numbers defined below. Although droplet impact
on solid surfaces or on liquid films show similar output, the physical
mechanisms leading to these effects often have different origins.  For
droplet impact on solids, the surface properties play an important role,
through its roughness and the contact line dynamics for instance.
A remarkable discovery has been done recently: the surrounding gas (usually
air) also plays a crucial role in splash
formation~\citep{Xu05}, and understanding in detail the influence of the gas still
remains a challenge~\citep{Mandre12,Gordillo14,klaseboer2014universal}. In
particular, the formation of a thin air layer at the instant of impact
smoothes the singularity expected in the absence of any gas and
thus ``cushions'' the impact. It also leads to the entrapment of an air 
bubble~\citep{thoroddsen03,chandra03,thoroddsen05,Mandre09,DJ11},

In this paper we focus on droplet impact on a thin liquid film where
splashing and spreading depend mostly on the balance, in the liquid as
well as in the gas, between
inertial and viscous forces~\citep{StHa81,MST95,YW95,JZ03}. In the literature about impacts
on liquid films, the effect of the surrounding air has not been shown as dramatically
as on solid surfaces, although a systematic study of its influence is still
lacking. For instance, it has  often been noticed experimentally and
numerically that air bubbles were entrapped by the impact
dynamics~\citep{thoroddsen03,Thor12}, and the interplay between these
bubbles and the ejecta sheet still needs to be elucidated.  In 2003 two of us~\citep[][which we further refer to as JZ03]{JZ03}
have proposed that the splashing/spreading transition observed in
experiments and in numerics was controlled by the capillary-inertia
balance within the ejecta sheet. The thickness in this theory was selected by a
viscous boundary layer. 
In such a model, scaling laws
for the jet thickness and velocity were deduced but the gas dynamics
was totally neglected and the existence of an entrapped bubble was not
considered at all.

The goal of this paper is therefore to determine the properties of the
ejecta sheet using high resolution numerical simulations of axisymetric 
incompressible newtonian two-phase flow, in order to exhibit the relevant physical
mechanisms at the heart of the prompt splash in this framework.

\section{The general problem}\label{diman}
\label{theory}
\subsection{Geometry and dimensional analysis}
\begin{figure}
\centering \includegraphics[width=8. cm]{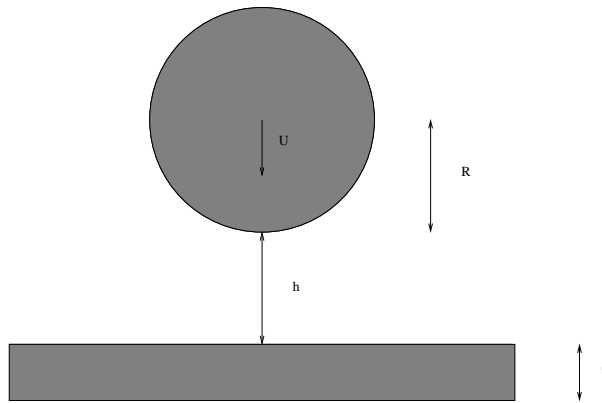}
\caption{Sketch of the numerical simulation, a spherical droplet of radius $R$ and velocity $U_0$ is considered at a distance $h_0$ from a liquid film of thickness $e$.}
\label{sketch}
\end{figure}

We consider a droplet of diameter $D$ impacting on a
thin liquid film of thickness $e$ with a velocity $U_0$ normal to the film
interface. Both liquid and gas have densities $\rho_l$ and
$\rho_g$, and dynamical viscosities $\mu_l$ and $\mu_g$
respectively. The surface tension is $\sigma$. 
In experiments the droplet is typically produced from some
height $H$ above the film and falls under gravity $g$. In our case 
the simulation starts
with a small initial air gap $h_0$ between the film and the droplet
and a velocity $U_0$, as shown
on Figure~\ref{sketch}. The droplet is assumed to be spherical. 
In order for this assumption to be valid we need a) to have as little effect 
of the air flow on the droplet shape as possible. This should be
verified if the gas Weber number $ \We_g = {\rho_g U_0^2 D}/{\sigma}$
is small or in a viscous regime and b) to 
assume that the oscillations of the droplet shape caused
by the droplet release mechanism \citep{Wang} are
as small as possible. 
The Froude 
number $\Fr =U_0^2/(g D)$ that quantifies the influence of the gravity during the impact is taken constant and high ($\Fr=800$) for all the simulations, indicating that gravity has only a small effect on the dynamics. 
%(However, in the numerical simulations that are shown here, gravity is present.) 
We will restrict this study 
to large liquid Weber numbers  $\We =  {\rho_l U_0^2 D}/{\sigma}$.

The problem is then mostly characterized by the dimensionless numbers
$$ 
\re= \frac{\rho_l U_0 D}{\mu_l}, \;\;\; \;\;\; \St=\frac{\mu_g}{\rho_l U_0 D}  \;\;  {\rm and}  \;\; 
\alpha = \frac{\rho_g}{\rho_l}
$$
which are the liquid Reynolds and Stokes numbers of the impact, and the density ratio. These
numbers compare the droplet inertia with viscous effects in the liquid and gas 
respectively, and compare the liquid to the gas inertia. We note that $\rho_l/\rho_g$ has to be
very large if one wants simultaneously to have $\We_g \ll 1$ and $\We \gg 1$. 
An additional dimensionless number is the aspect ratio
between the liquid film and the droplet $e/D$ which we keep relatively small and constant in this study. We expect the 
initial gas layer aspect ratio $h_0/D$ to be irrelevant if the conditions
described above ($\Fr \gg 1, \We_g < 1$) are satisfied and $h_0$ is larger than
the characteristic thickness $h_b$ defined below. 
Compressibility effects are characterized  by the Mach numbers 
$\Ma=U/a_{l,g}$, where $a_l$ (resp. $a_g$) is the speed of sound
in the liquid (resp. gas) and in all our simulations, these Mach numbers remained small enough so that
compressibility effects could be neglected~\citep{LeFi83}.

We note that the axisymmetric flow assumption is not valid
when digitations and splash droplets form. However, at short time, before these instabilities can
develop and particularly before the jet is created, we can consider this assumption correct {\bf citer et s'inspirer des resultats experimentaux}.
Otherwise the
general 3D problem remains a grand numerical challenge
because of the large range of scales involved~\citep{Rieber98,Gue98}. 
Despite some recent numerical results~\citep{Fuster}
realistic 3D numerical simulations of droplet impact at short times
are yet hard to attain. 

Moreover a solid basis for the analysis 
of the scaling of 3D flow may only be attained when the 
scaling of 2D flow has been uncovered. 
We thus postpone a detailed 3D study of droplet impact to future work.

\subsection{Scaling analysis}

We analyze now the different mechanisms at play during droplet
impact using simple scaling arguments. Recall that surface tension and gravity can be
neglected, and in a first step, we will consider also that the
surrounding gas has negligible effects. We then quite naturally define $t=0$ as the time at which an
undeformed, spherical
droplet at uniform velocity 
would touch the undeformed, planar liquid surface. With this definition of the time, the
origin initial time is $t_0 =-{h_0}/{U_0}.$

We will now use an important geometrical argument first suggested  by~\citet{Wagner32}:
considering the intersection of the falling
sphere with the impacted film, it is straightforward to define the
vertical lengthscale as $\ell_z = U_0t$ and the horizontal one 
$r_g=\sqrt{D \ell_z}=\sqrt{DU_0 t}$. 
These apparently simple scalings arising purely from geometry are in
fact very robust and relevant to the description of impact at short times:
for instance, it has been shown that $r_g(t)$ gives a correct estimate
of the so-called  {\it spreading} radius defined as the radius where the
pressure is maximal. Remarkably, the geometrical
velocity of this intersection
\begin{equation}
v_g(t)= \frac{{\rm d}r_g}{{\rm d}t}=\frac12\sqrt{\frac{DU_0}{t}}, \label{vgeq}
\end{equation}
diverges at $t=0$, questioning the incompressible assumption. However, although formally such a
geometrical velocity diverges, fluid velocities remain much smaller and compressibility can be safely 
neglected for small Mach numbers. 
To make the article self-contained, we will now recall rapidly
the results obtained by JZ03. The key point is
the numerical observation that the pressure field and the
velocity field are perturbed over the length scale $r_g(t)$ so that a
kind of inner-outer asymptotic analysis can be performed, in which 
the flow is uniform at scales larger that $r_g$, potential at scales
of order $r_g$ and viscous at scales much smaller than $r_g$ (a more rigorous asymptotic analysis
has been developed later in \cite{HoOc05}). 
In this analysis the ejecta speed is
obtained using a mass conservation argument between the impacting
droplet and the ejected sheet, assuming that the thickness of the jet
is selected by a viscous length. More precisely, one can compute in
this framework first the mass flux $F_m$ from the falling undeformed
sphere through the undeformed film surface
\begin{equation}
F_m(t) \sim \rho_l \pi r_g^2 U_0. 
\label{Fmass}
\end{equation}
This flux can be absorbed either by surface deformation of the droplet
and of the film or by the formation of an ejecta sheet. In JZ03, we have
assumed that the thickness of such a sheet or {\it jet} is given by a viscous boundary
layer formed at the basis of the jet leading to a viscous length scale
\begin{equation}
e_j(t) \sim \sqrt{\frac{\mu_l t}{\rho_l}}. 
\label{jetthick}
\end{equation}
Then conservation of the volume flux through this jet implies that the jet velocity $U_j$ has to satisfy
\begin{equation}
U_j \sim  \sqrt{Re} U_0.
\label{jetvel}
\end{equation}
Remarkably, this gives a nonlinear relationship between the jet and
the impacting droplet velocity since then $U_j \propto U_0^{3/2}$. Such a
law has obviously some physical restrictions: first of all, the flux formula
(\ref{Fmass}) is valid only for $t\ll R/(2U_0)$ since our whole
analysis is for short times, and makes no sense for times 
of order $D/U_0$. Furthermore, the jet velocity has to be 
larger than the geometrical velocity $v_g$. Indeed if 
$U_j < v_g(t)$ one would expect the ejecta sheet to be overrun by the falling
droplet.  This condition together with (\ref{vgeq}) yields a ``geometric'' limiting time $t_g$
\begin{equation}
t > t_g \sim \frac{1}{\re} \frac{D}{U_0}. 
\label{tgcond}
\end{equation}
When the ejecta forms, a bulge or rim appears at its tip
according to the mechanism of~\cite{Taylor59c} and \cite{Culick60}. 
This rim moves backwards at the Taylor-Culick velocity 
\begin{equation}
 v_{TC}= \sqrt{\frac{2 \sigma}{\rho_l e_j}}.
\label{TCvel}
\end{equation}
where $e_j$ is the thickness of the ejecta sheet. 
The ejecta sheet cannot form if its velocity is smaller than
the Taylor-Culick velocity constructed with the thickness $e_j$ of the ejecta sheet.
We thus obtain that the ejecta can form only when $U_j > v_{TC}$ which from equations
(\ref{jetvel}, \ref{jetthick}, \ref{TCvel}) yields
\begin{equation}
 t > t_{TC}= \frac{2}{\We^2 \re} \frac{D}{U_0}. \label{tccond}
\end{equation}
Both conditions (\ref{tgcond}) and (\ref{tccond}) must be satisfied at short times 
$t \ll D/U_0$ because as stated above the whole theory does not make sense
for larger times. Then we must have ${\rm max}(t_{TC},t_g) \ll D/U_0$ which yields 
the condition
$${\rm min}( \We^2 \re,\re)  \gg 1. $$
We restrict the present study to the dynamics where splashing is always present.
More precisely since we consider situations such that $\We \gg 1$ and $\re \gg 1$,
then of the two conditions for splashing given above,  
$t>t_g$ is always more restrictive than $t>t_{TC}$. Therefore, in our configuration, the jet appears
only when its velocity is bigger than the geometrical one.

Finally, the inertial pressure of the impact $P_{imp}$ can be
computed using the rate of change of vertical momentum in the droplet,
following JZ03:
\begin{equation}
 P_{imp} \pi r_g^2 \sim \frac{2\pi}{3} \rho_l r_g^3 U_0  \frac{{\rm d}r_g}{{\rm d}t} .\label{Pimp}
\end{equation}
In this equation, the vertical momentum in the droplet is affected only in a half-sphere of radius $r_g$.
Eq. (\ref{Pimp}) gives the impact pressure 
\begin{equation}
 P_{imp} \sim 2 \rho_l \frac{{\rm d}r_b}{{\rm d}t} U_0
\end{equation}
which leads to 
\begin{equation}
 P_{imp} \sim \sqrt{\frac{D}{U_0t}} \rho_l U^2_0 \label{joss_p_imp1}
\end{equation}
as observed in numerical simulations (see JZ03).
Note that a detailed analysis of the potential flow
for a droplet falling on a solid surface has been performed by
\citet{philippi15}. The reasoning in the latter paper may be straightforwardly
transposed to the impact on a liquid surface to obtain for the pressure field 
in the neighborhood of $z=0$
\begin{equation}
 p(r,t) \sim \frac {3 \rho_l U^2_0 } {\pi \sqrt{\frac{3 t U_0}{R} - \frac{r^2}{R^2}}} \label{cop_p_imp}
\end{equation}
which is very similar at $r=0$ to the scaling in (\ref{joss_p_imp1}). However
the pressure field of (\ref{cop_p_imp}) has an additional singularity for $t>0$, not
predicted by Equation (\ref{joss_p_imp1})  at 
$r = \pm \sqrt{3RU_0t}$. This singularity is indeed observed in our 
numerical simulations as well as in JZ03 and in \citet{philippi15,DJ11}.

In the theory above, contact occurs at $t=0$, 
the vertical length scales are $U_0 t$ and $e_j$, the flow pressure and the geometric velocity are singular with an infinite limit at $t=0$.
However, the effect of the gas layer was not taken into account so far.
When instead the gas layer is taken into account, the above analysis is an approximation
valid at length scales $\ell \gg h_b$ where $h_b$ is the thickness of the 
gas layer. The scale of the impact pressure is thus
\begin{equation}
 P_{imp} \sim \sqrt{\frac{D}{h_b}} \rho_l U^2_0 \label{zal_p_imp1}
\end{equation}
The singularity of velocity and pressure is regularized and
it can be said that the gas ``cushions'' the shock of the impact.

Moreover it is observed in 
experiments by \cite{thoroddsen03} and in numerical
simulations~\citep{JZ03,chandra03,Korob08} that contact does
not occur on the symmetry axis $r=0$ but on a circle of radius $r_b$
so that a bubble is entrapped, as observed also for droplet
impact on solid surfaces by \citet{thoroddsen05,kol12}.
As before, horizontal and vertical length scales are related at short times by 
$l_z \sim l_r^2/D$. Thus the thickness $h_b$ of the gas layer or bubble 
at the time of contact relates to the contact radius $r_b$ by  
\begin{equation} 
h_b =r_b^2/D. \label{hrDrel}
\end{equation}
These short-time asymptotics have to match the initial conditions
at negative time $t=-t_0$. Let $z_+(r,t)$, $z_-(r,t)$ be the positions
of the drop and film surfaces respectively,
and $h(r,t) = z_+(r,t) - z_-(r,t)$ be the thickness of the gas layer.
To fix ideas, let us consider initial conditions such that
$z_-(r,-t_0) = 0$, $h(r,-t_0) = z_+(r,-t_0)$  close to the impact time
so that on the axis $h(0,-t_0) = h_0 \ll D$. Then the gas layer is 
thin, there is a separation of horizontal and vertical length scales
so that a lubrication approximation is valid over distances
$l_r \sim \sqrt{h(0,-t) D}$.
As long as the lubrication pressure (estimated below) thus obtained
in the gas layer is much smaller than
the impact pressure $P_{imp}$ the liquid advances almost undeformed
while expelling the gas.
In this regime
\begin{equation}
\partial_t h (0,t) = - U_0. \label{initialdhdt}
\end{equation}
When the lubrication 
pressure becomes large enough to deform the liquid and 
slow down the thinning of the gas layer, 
the time is of order of a so called air-cushioning time scale $t_b$ and the thickness
reaches the air cushion length scale $h_b$.
Matching with the initial velocity, eq. (\ref{initialdhdt})
then gives the relationship $h_b = U_0 t_b$ 
 which together with the separation of scales condition
(\ref{hrDrel}) links the time and position of 
the contact  through $t_b \sim r_b^2/(DU_0)$.

In order to determine these air cushioning space and time scales, we 
find the dominant balance in the lubrication
equation, following in part recent works on impacts on solid
surfaces~\citep{Korob08,Mandre09,DJ11,klaseboer2014universal}. 
Our theory starts from the incompressible lubrication equation in cylindrical geometry:
\begin{equation}
\partial_t h=\frac{1}{12 \mu_g r}\partial_r(rh^3 \partial_r P)
\label{eq:lubr}
\end{equation}
where $P(r,t)$ is the pressure in the gas layer.
The factor $1/12$ in front of the lubrication pressure comes from the
Poiseuille velocity profile valid for laminar flows, obtained with a zero radial velocity
at  $z_-$ and $z_+$, which is assumed because of the small horizontal velocity in the liquid before
splashing. Using the above geometrical argument $ h \sim \sqrt{Dr}$ for the pressure term and
$\partial_t h \sim -U_0$, we obtain the following scaling for the lubrication pressure $P_b$ in the gas film of thickness $h$:
$$ P_b\sim \frac{3 \mu_g U_0 D}{h^2},$$
The usual lubrication scaling for the bubble entrapment is then obtained
by writing that during this ``cushioning phase'', the air pressure for $h\sim h_b$  balances the 
impact pressure, that is $P_b \sim P_{imp}$ yielding:

$$ P_b \sim \frac{3 \mu_g U_0 D}{h_b^2} \sim P_{imp} \sim \rho_l U_0^2 \sqrt{\frac{D}{U_0t_b}} \sim \rho_l U_0^2 \sqrt{\frac{D}{h_b}}.$$
Here we have used the relation $h_b=U_0 t_b$, where $t_b$ is the time of bubble entrapment.
This relation gives the following scaling for the bubble entrapment:
\begin{equation}
h_b \sim \St^{2/3} D\, , \;\; r_b \sim \St^{1/3} D \;\; , \;\; t_b \sim \frac{h_b}{U_0} \sim \St^{2/3} \frac{D}{U_0}
\;\; {\rm and} \;\; P_b \sim \frac{3 \mu_g U_0 D}{h_b^2} \sim 3 \rho_l U_0^2 \St^{-1/3}. 
\label{lubscal}
\end{equation}

The cushioning phase starts when
$t$ is {\em negative} and of order $-t_b$ and ends 
when first contact occurs at a {\em positive} time $t_b$. 
This leads to two remarks: one is that the time of ``cushioning'' $t_b$ is both 
the time scale of the {\em duration} of this phase, and the time coordinate
of the {\em two instants} of the beginning of the cushioning phase and the end of
it at the first contact. We do not have strong arguments or data to show that this
two instants are symmetric around $t=0$. 
However, interestingly, our numerical simulations show that a kind of droplet/film
symmetry holds, so that
at $t=0$, we have to a high degree of accuracy $z_{+}(0,0) = - z_-(0,0)$ (see Figure \ref{z_t} below).

In this approach, in agreement with previous works ~\citep{Mandre09,DJ11}, gas inertia effects have 
been totally neglected but such an assumption is questionable, as suggested in a recent study~\citep{Gordillo14}. We propose here to take into account inertial pressure losses for the 
pressure in the gas layer using simple scaling arguments. Implementing inertial effects in thin film 
equation remains a difficult problem which is still a question of scientific debate~\citep{Luchini2010} and here
we will only estimate scaling laws for the contribution of inertia. 

The variation of the film height is given by the divergence of the horizontal gas flux in the layer
$$  
\partial_t h= - \frac1r \partial_r ( r h \bar u ) ,
$$
where $\bar u$ is the averaged horizontal velocity between $z_-$ and $z_+$. 
We can determine the scale $u_b$ for $\bar u(r,t)$  by considering
the momentum balance in a thin gas layer
$$
\partial_t \bar u  + A \bar u  \partial_r \bar u  = - \frac 1 \rho_g \partial_r p - 12 \mu_g 
\bar u /h^2 - K {\bar u}^2/h
$$
where $A$ is a constant depending on the profile of the flow in the gas
layer and $K$ is a constant characterizing turbulent friction. This equation
will hold if the flow remains thin ($h\ll D$) and does not separate. 
Since in incompressible flow
pressure is defined up to a constant and the pressure at the exit of the thin
gas layer flow (taken here for $r \sim r_g(t)$) is the pressure at infinity, it
is convenient to set this pressure at the exit to zero. 
Then $P_b$ equals the pressure difference and can be estimated at the bubble entrapment yielding
\begin{equation}
P_b = 12 \mu_g r_b u_b/h_b^{2} + C_1 \rho_g u_b^2 r_b/ h_b  
+ C_2 \rho_g u_b^2.  \label{modiflub}
\end{equation}
 Here, we have taken for the first (dominant) term the lubrication pressure already computed above.
 The second  and third
terms result from two kinds of inertial effects, the turbulent friction term and a possible singular head loss due to flow separation.
It is readily seen that
the ratio between the first two terms is the local Reynolds number of the gas layer
$\rho_g u_b h_b/\mu_g$. 
The third term is a singular head loss. 
The constant
$C_1$ and $C_2$ depend on the precise geometry of the flow and are difficult
to estimate. However, it can be seen that the singular head loss is much smaller for our problem by a 
factor $h/r_g = (D U_0/t_b)^{-1/2}$ than the turbulent friction term so that we will neglect the singular head loss in the following developments. 

The film pressure may be finally obtained  by estimating the horizontal
velocity scale as $u_b = r_b/t_b$ which together with 
eq. (\ref{hrDrel})
 yields
\begin{equation}
P_{b} \sim \frac{12 \mu_g U_0 D}{h_b^{2}} + C_1 \rho_g U_0^2 \left(\frac{D}{h_b}\right)^{3/2}. 
\end{equation}

Remarkably, the neglected singular head loss term would give an additional contribution in the form $C_2 \rho_g U_0^2 D/h_b$.

Finally, equating $P_{b}$ and $P_{imp}$ yields now
an implicit equation for $h_b$ 
\begin{equation}
 \rho_l U^2_0 \left(\frac{D}{h_b}\right)^{1/2} =  \frac{12 \mu_g U_0 D}{h_b^{2}} 
+  C_1 \rho_g U_0^2 \left(\frac{D}{h_b}\right)^{3/2}
%+  C_2 \rho_g U_0^2 \frac{D}{h_b}
\end{equation}
which can be written in terms of the dimensionless variables 
$\hat{h}_b = h_b/D$, $\St$ and $\alpha$
\begin{equation}
12 \, \St =  \hat{h}_b^{3/2}  - {C_1\alpha}\hat{h}_b^{1/2}  \label{hbtheory}
\end{equation}
The above equation is cubic in $\xi = \hat{h}_b^{1/2}$ 
and its solution gives the dimensionless height of the film as
a function of $\St$ and $\alpha$ with two asymptotic regimes separated by a critical 
Stokes number 
\begin{equation}
\St_c =\frac 1 {12} (C_1 \alpha)^{3/2}.  \label{Stcr}
\end{equation}
The first regime, for 
$\St \gg \St_c$ corresponds to the case computed above with lubrication only (\ref{lubscal}) and is of the form
$$ 
\hat{h}_b \sim {\rm St}^{2/3}
$$
as also stated by \citet{Mandre09,Mani10}.
In the other regime, 
for $\St \ll \St_c$  we get
\begin{equation}
\hat{h}_b \sim C_1 \alpha. \label{hcr}
\end{equation}
The estimates for the time at which the bubble is entrapped results
from the estimates for $h_b$ through $t_b = h_b/U_0$. For $\St \gg \St_c$ we recover eq. (\ref{lubscal})
\begin{equation}
\hat{t}_b  \sim \St^{2/3}
\label{eqtb}
\end{equation}
 where the dimensionless time is noted $\hat t = U_0 t / D$,
and all the other scalings are obtained straightforwardly from the scalings for the height
$\hat{h}_b$ given above. 
Similarly from Equation (\ref{joss_p_imp1}) the pressure in and on top of the gas layer,
neglecting surface tension effects is (eq. \ref{lubscal}):
\begin{equation}
 P_{imp, max} \sim \rho_l U^2_0 \St^{-1/3} \label{max_p_imp}
\end{equation}
This theory will now serve as a framework to interpret the 
numerical simulations reported below. The main prediction is that
air delays contact by a time of order $t_b$ and that a bubble of typical radius $r_b$ is entrapped. 

The above considerations however do not say at what time $t_j$ the liquid sheet is ejected. 
It sets contraints on the air cushioning effect and we can only turn to numerical simulations
to see how the air layer dynamics interacts with jet formation.

To conclude this scaling analysis, it is interesting to notice that an alternative theory has been 
proposed recently~\citep{klaseboer2014universal}. There a different scaling for the entrapment 
has been obtained (leading to $ h_b \sim St^{1/2} D$ instead of $h_b \sim St^{2/3} D$) based on the balance between the lubrication pressure in the gas and the Bernoulli pression in the drop $\rho_l U_0^2$. Although the pressure amplitude in numerical simulations of drop impact has been shown to 
obey the singular law~(\ref{cop_p_imp}),
experimental studies have yet to distinguish between these two predictions. 

\section{Numerical method}\label{num}

In our continuum-mechanics modelling approach, fluid dynamics is
Newtonian, incompressible, with constant surface tension. In the
``one-fluid approach'' \citep{Tryggvason11} one considers a single
fluid with variable viscosity and density, and a singular surface
tension force, yielding the Navier-Stokes equations that read:
\begin{equation}
 \rho {\partial {\bf u} \over \partial t}+\rho \nabla \cdot  {\bf u} {\bf u} 
=-\nabla p + \nabla \cdot \mu [ \nabla {\bf u} + (\nabla {\bf u})^T ] + \sigma \kappa{\bf n} \delta_s .
\label{m8}
\end{equation}
\begin{equation}
{\rm div} ({\bf u})=0 \label{incomp}
\end{equation}
where ${\bf u}$, $p$ is the pressure, ${\bf n}$ denotes the unit
normal to the interface and $\delta_s$ is the two-dimensional Dirac
distribution restricted to the interface, $\rho(x)$ and $\mu(x)$ are
the space-dependent fluid densities and viscosities equal to their
respective values $\rho_{l,g}$ and $\mu_{l,g}$ in each phase.  This
set of equations can be written using dimensionless variables,
rescaling lengths by $D$, velocities by $U_0$, times by $D/U_0$, densities
by $\rho_l$ and pressures by $\rho_l U_0^2$ so that the Navier-Stokes
equations become:
\begin{equation}
\rho \frac{\partial {\bf u}}{\partial t}+ \rho  \nabla \cdot  {\bf u} {\bf u}  
 = -  {\bf \nabla p} +  {\bf \nabla} \cdot \frac{\mu}{\re} [ \nabla {\bf u} + (\nabla {\bf u})^T ]
+ \frac{\kappa}{\We}{\bf n} \delta_s .
\label{NSadim}
\end{equation}
where now $\rho=\mu=1$ in the liquid phase while $\rho=\alpha=\rho_g/\rho_l$
and $\mu=\mu_g/\mu_l$ in the gas. 
Equation~(\ref{NSadim}) is solved using the methods described in 
\citep{Pop03,GerrisVOF,Lagree:2011bq,Tryggvason11}, that is by discretizing the fields on 
an adaptive quadtree grid, using a projection method for the pressure,
 the time stepping and the incompressibility condition. The advection of
the velocity fields is performed using the second-order Bell-Collela-Glaz scheme, 
and momentum diffusion is treated partially implicitly. The interface is tracked
using a Volume of Fluid (VOF) method with a Mixed Youngs-Centered Scheme  \citep{Tryggvason11} 
for the determination of the normal vector
and a Lagrangian-Explicit scheme for VOF advection. Curvature is computed using the
height-function method. Surface tension is computed from curvature 
by a well-balanced Continuous-Surface-Force method. 
Density and viscosity are computed from the VOF fraction $C$ 
by an arithmetic mean. This arithmetic mean is followed by 
three steps of iteration of an elementary filtering. 
This whole set of methods is programmed either in the
{\sc Gerris} flow solver \citep{Gerris}, or in the {\em Gerris scripts}
that were designed to launch these computations.

Four refinement criteria are used as follows 1) the local value of the vorticity,
2) the presence of the interface as measured by the value of the gradient
of the VOF ``color function'' 3) a measure of the error in the discretisation of the
various fields based on an {\em a posteriori} error estimate of a given field as a cost function for adaptation. This {\em a posteriori} error is estimated by computing the norm of the Hessian matrix of
the components of the velocity field, estimated using third-order-accurate discretisation operators,
4) when near the interface, the curvature is used as the adaptation criterion. 
To measure the degree of refinement so obtained, recall that 
on a quadtree grid, a level of refinement $n$ means that the
grid cell is $2^{n}$ times smaller than the reference domain or ``box''. 
When adaptively refining, a predefined maximum level $n_0$ is used
for the adaptation on curvature (4), moreover adaptation on vorticity (1)
and on the error (3) may lead to a maximum level of refinement $n_0-1$ and finally
cells near the interface (2) are always refined to level $n_0 -2$ at the least. 
Note that criterion (3) is generally more efficient than (1) so the latter could
have been dropped altogether. 
 
\section{Results of simulations}

\subsection{Impact dynamics}
\label{results}

We perform series of simulations with parameters set as in Table~\ref{allparams}. For water-like fluids, 
these constant Weber and Froude numbers correspond to a drop of radius $1$ mm falling at velocity 
$4$ $m\cdot s^{-1}$. The numerical simulations are performed for different liquid and gas viscosities 
characterized by the Reynolds and Stokes numbers varying from $400$ to $16000$ and from 
$5.65 \cdot 10^{-7}$ to $2.26 \cdot 10^{-5}$ respectively. For a $2$ mm diameter drop impacting at 
$4\; {\rm m} \cdot {\rm s}^{-1}$, it would typically cover the range between
one eighth to twenty times the water viscosity, and one fourth to ten times the air viscosity. In 
particular, we have done simulations for three Stokes numbers 
($2.26 \cdot 10^{-6}$, $9.05 \cdot 10^{-6}$ and 
$2.26 \cdot 10^{-5}$) a large range of Reynolds numbers ($400$, $600$, $800$, $1000$, $2000$, 
$4000$, $8000$ and $16000$) that will be used to analyse the dynamical properties of the impact.

\begin{table}
\begin{center}
\caption{Dimensionless values of the parameters for all simulations reported} \label{allparams}
\begin{tabular}{ccccc}
\hline
 \We & \Fr & $\rho_l/\rho_g$ & $e/D$ & $h_0/D$ \\
\hline
500 &  815 & 826.4 & 0.2  & 1/30 \\
\hline 
\end{tabular}
\end{center}
\end{table}
In an initial phase the droplet falls undeformed until air cushioning effects set in. Then at some point in time
the jet forms and a reconnection of the interfaces on the droplet and film 
occurs. 
Figure~\ref{snap} shows a droplet impact with 
physical parameters approximating a glycerinated water droplet falling
in air. The main dimensional parameters are given in Table~\ref{tabledimsnap}, other
parameters approximate air at ambient temperature, leading to the dimensionless 
numbers $\re=2000$ and $\St=2.26\, 10^{-6}$ in complement to the dimensionless numbers of 
Table~\ref{allparams}.

\begin{table}
\begin{center}
\caption{Dimensional values of the main parameters for Figure \ref{snap}} \label{tabledimsnap}
\begin{tabular}{ccccc}
\hline
$U_0$  & $D$ & $\mu_l$ &  $\sigma$ & $e$\\
\hline
4 m s$^{-1}$& $2 \cdot 10^{-3}$m &  $4 \cdot 10^{-3}$kg m$^{-1}$s$^{-1}$ & $64 \cdot 10^{-3}$ 
 kg s$^{-2}$ & $4 \cdot 10^{-4}$ m\\
\hline 
\end{tabular}
\end{center}
\end{table}

\begin{figure}
\centering
a)\includegraphics[width=6.cm]{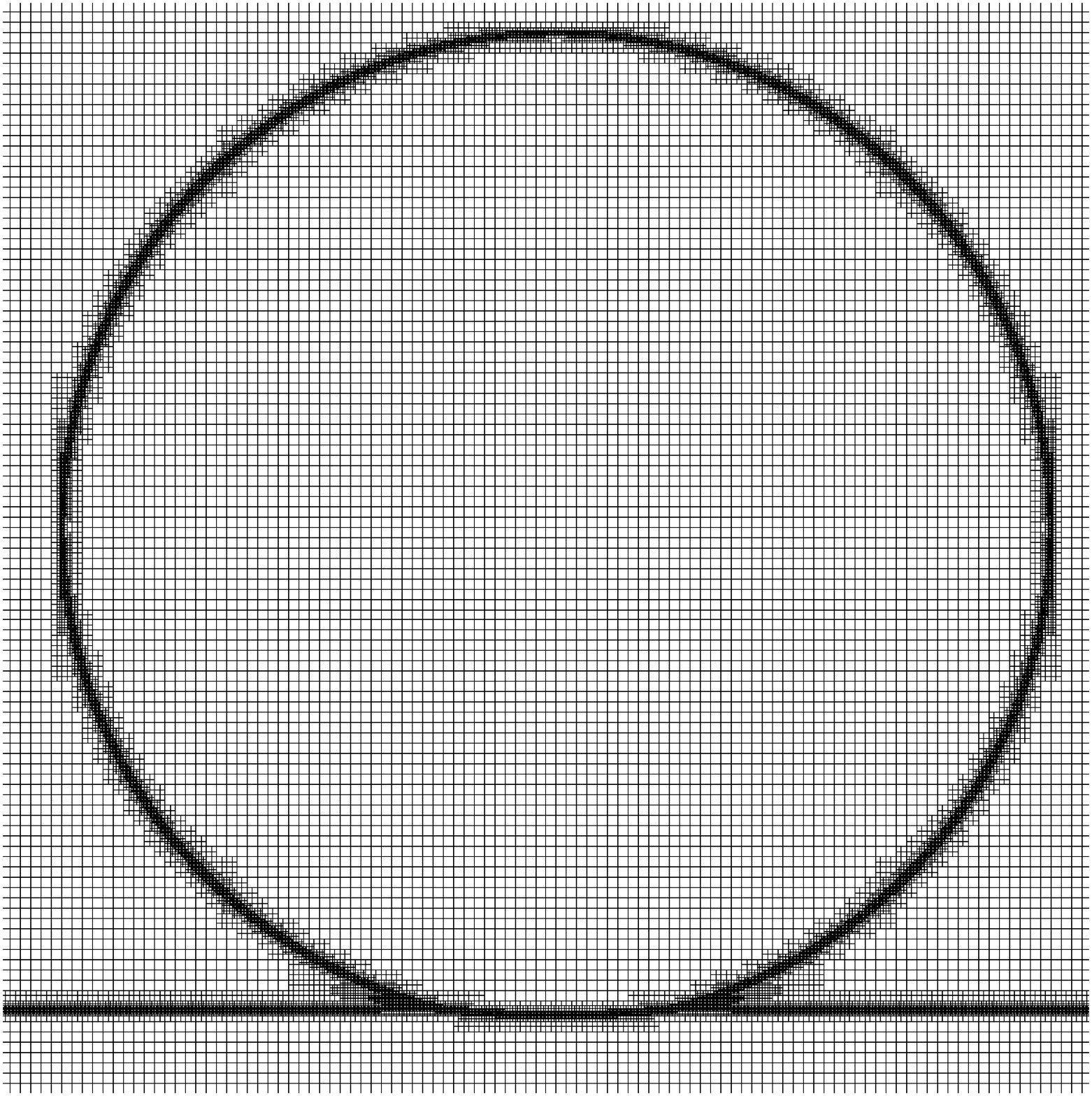}b)\includegraphics[width=6. cm]{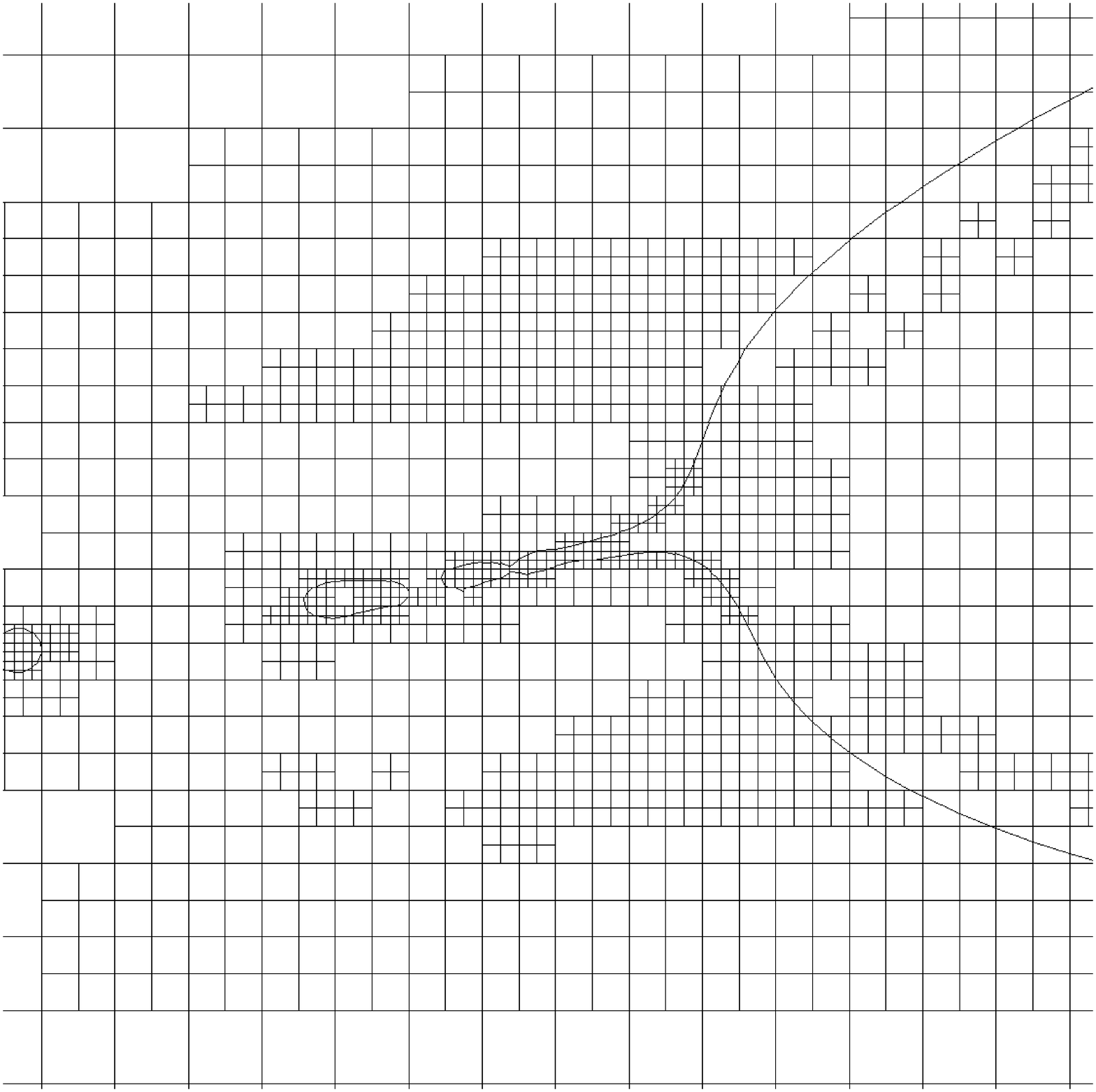}\\
\caption{The grid refinement used in a) Figure \ref{zoom}a and b) Figure \ref{zoom}d. Here, the viscosity is twice that of Table~\ref{tabledimsnap}, so that $\re=1000$ with a gas viscosity such that $\St=2.26\, 10^{-6}$.}
\label{zoomgrid}
\end{figure}

The grid is refined based on the four criteria above so that the smallest cell has
size $\Delta x = D/25000$. Figure \ref{zoomgrid} shows two views of the 
grid refinement for a case where the liquid viscosity is twice higher than in Table~\ref{tabledimsnap} the gas one ten times larger, so that $\re=1000$ and $\St=2.26\, 10^{-6}$, all the other parameters being the same than on figure~\ref{snap}

\begin{figure}
\centering
\includegraphics[width=10.cm]{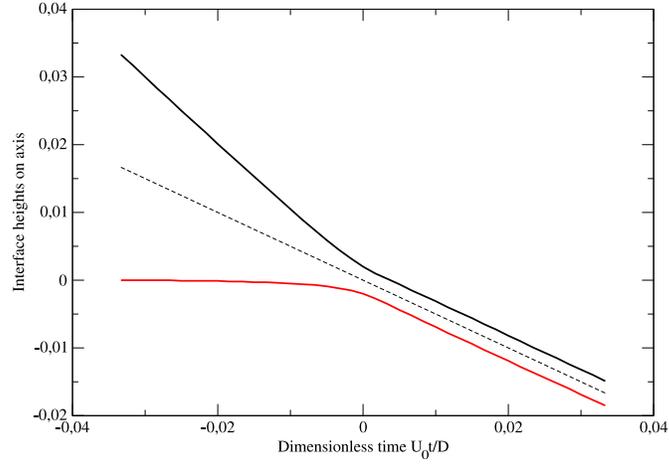}
\caption{The positions of the bottom of the droplet and the top of the film on the axis as a function of time
for $\St=2.26 \, 10^{-5}$, $\re = 1000$. The dashed line represents the mean position that decreases like $-0.5U_0 t$.}
\label{z_t}
\end{figure}

We have checked that higher $h_0/D$ does not change the
results significantly. This can also be verified 
from Figure~\ref{z_t} where it can be seen directly that the simulation
starts at a time $t_0 = - D/(30 U_0)$ that is much larger 
than the apparent time scale of the air cushioning effect. In fact, the velocity of the south pole of the droplet in unperturbed
until very short times around $t \sim -5 \cdot 10^{-3}D/U_0$. 

\begin{figure}
\centering
a)\includegraphics[width=6.cm]{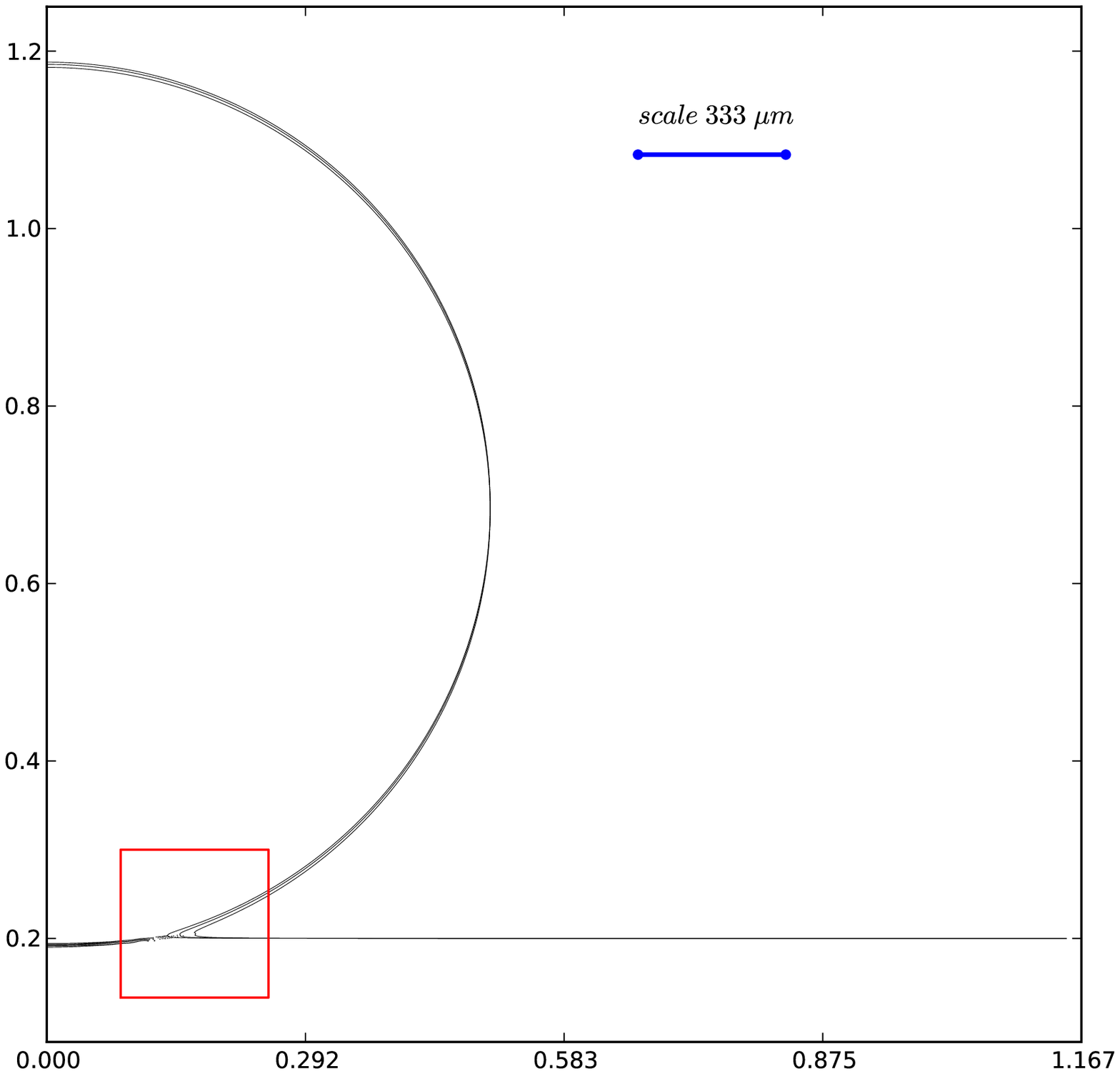} b)\includegraphics[width=6. cm]{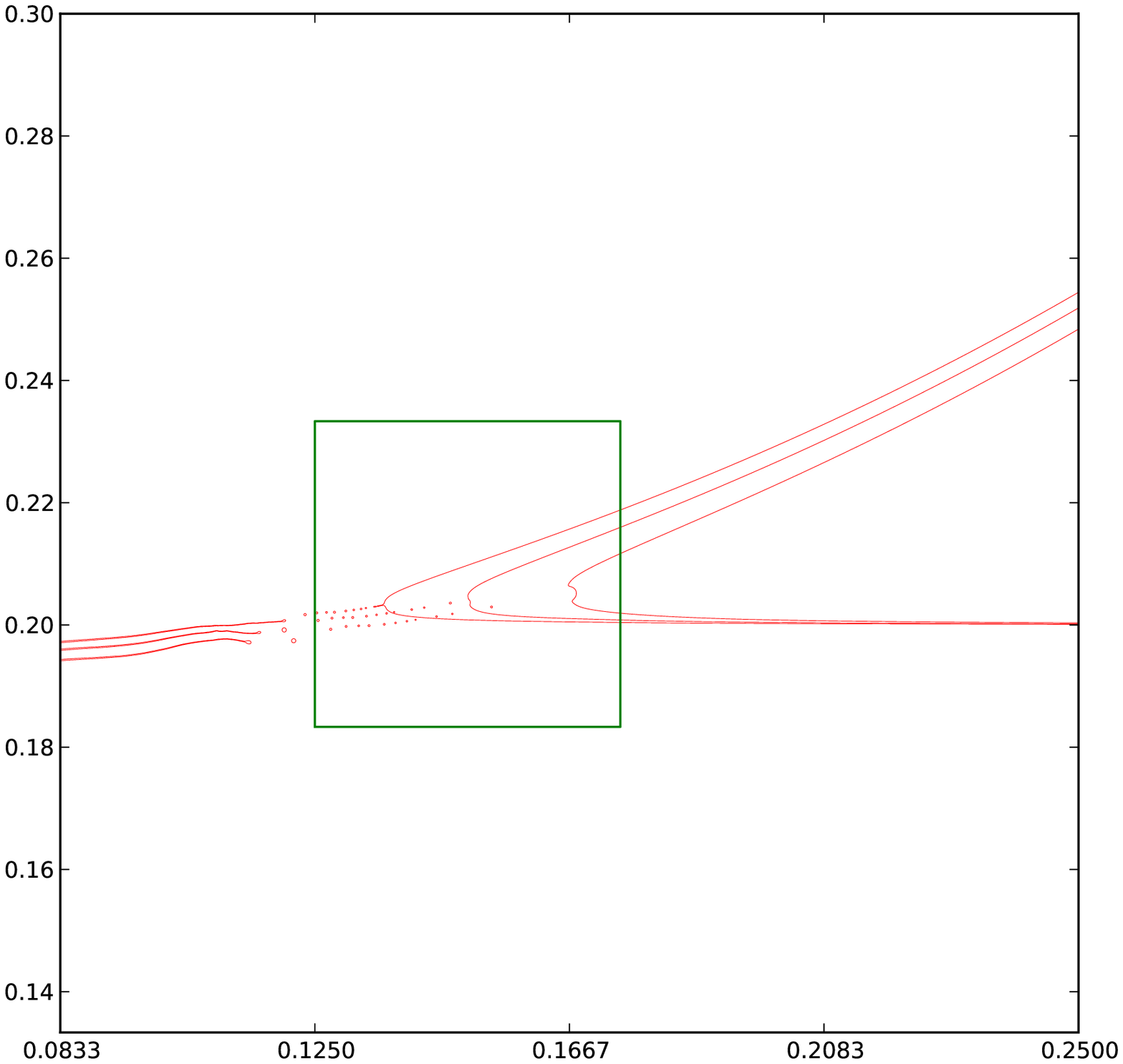}\\
c)\includegraphics[width=6. cm]{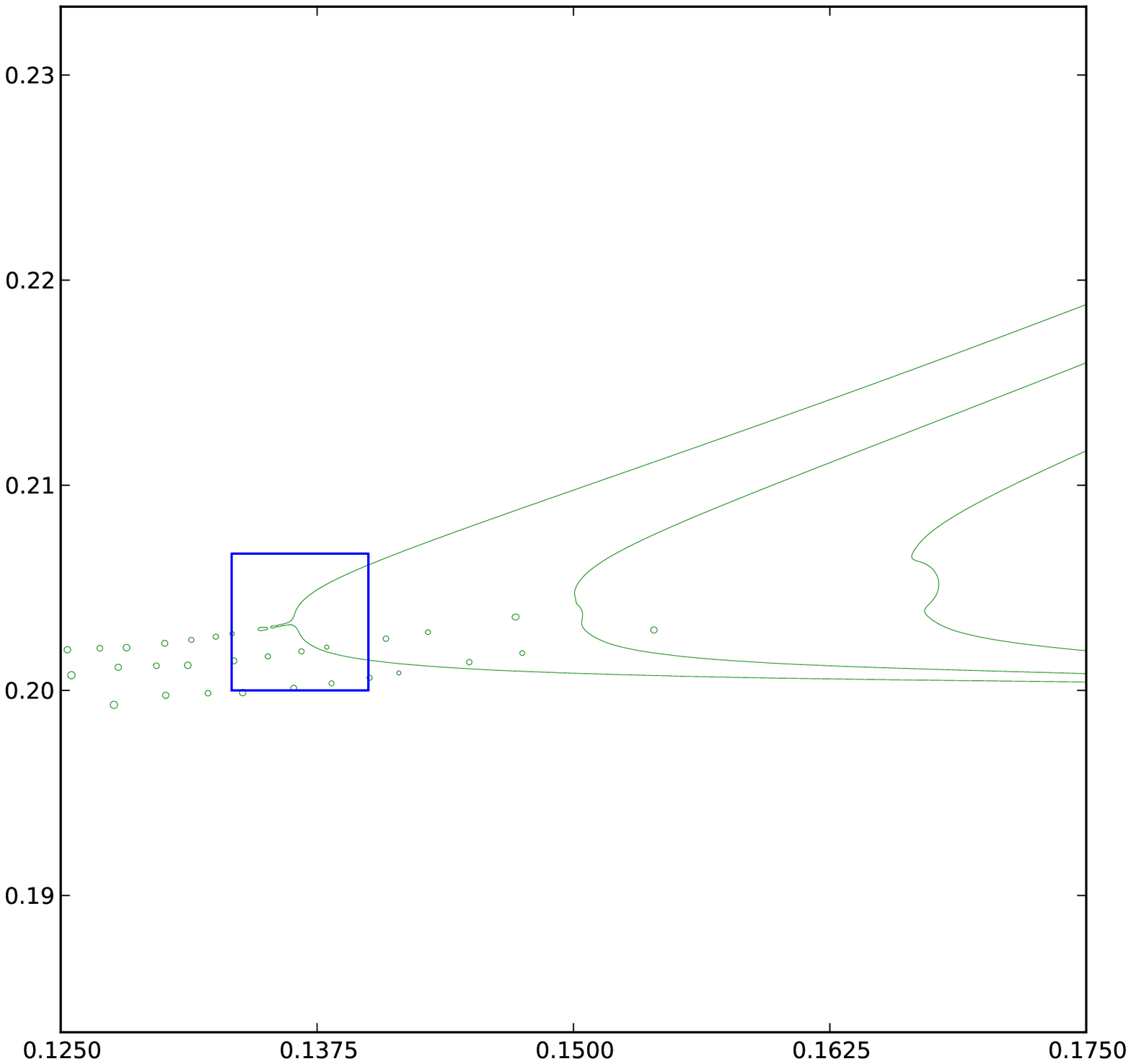} d)\includegraphics[width=6. cm]{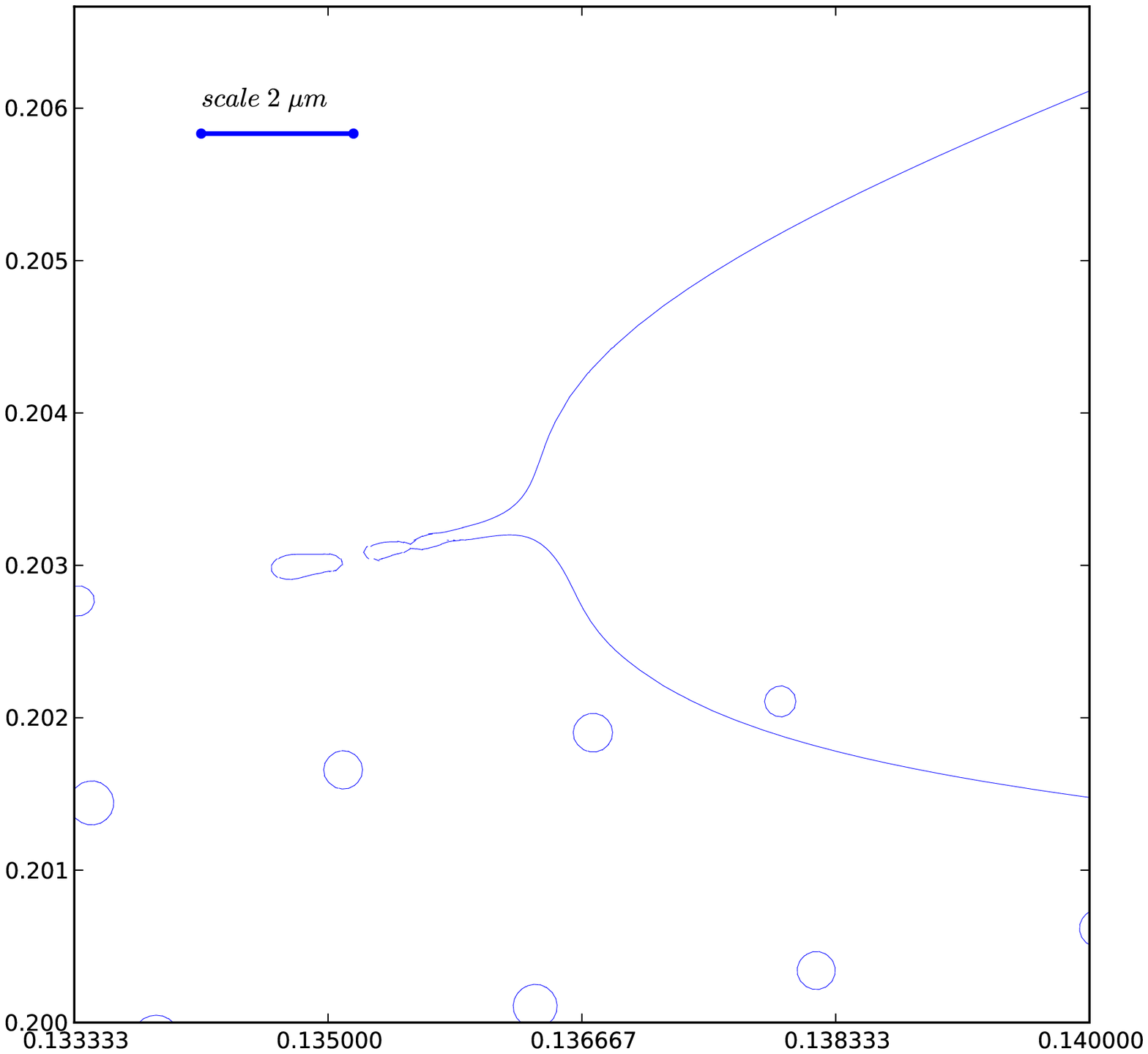}\\
\caption{Successive zooms of the interface at the same time $t=$ of impact for  $\re=1000$ and $\We=500$. a) shows the general view; b) shows the zoom of the interface corresponding to the square traced on figure a) and so on from b) to c) and c) to d). In the first and the latter figure, the physical scales are shown based on a $D=2$ mm diameter drop. The $2 \; \mu m$ scale shown in the latter figure corresponds in fact to twenty times the smallest mesh size of the numerical simulation.}
\label{zoom}
\end{figure}

In Figures~\ref{snap} and~\ref{z_t}, one can observe that a bubble
is indeed entrapped by the impact due to the cushioning of the gas
beneath the droplet. A very thin ejecta sheet is formed, followed by
the growth of a thicker corolla. These figures do not however give a full
account of the level of accuracy reached in the calculation, as
shown on Figure~\ref{zoom} where successive zooms of the interface are
shown around the instant when the droplet contacts the liquid film. 
The large range of scales between the droplet diameter and the 
small features in Figure~\ref{zoom}d is apparent,
Figure \ref{zoomgrid} showing the corresponding grid. 

\begin{figure}
\centering
a)\includegraphics[width=6.cm]{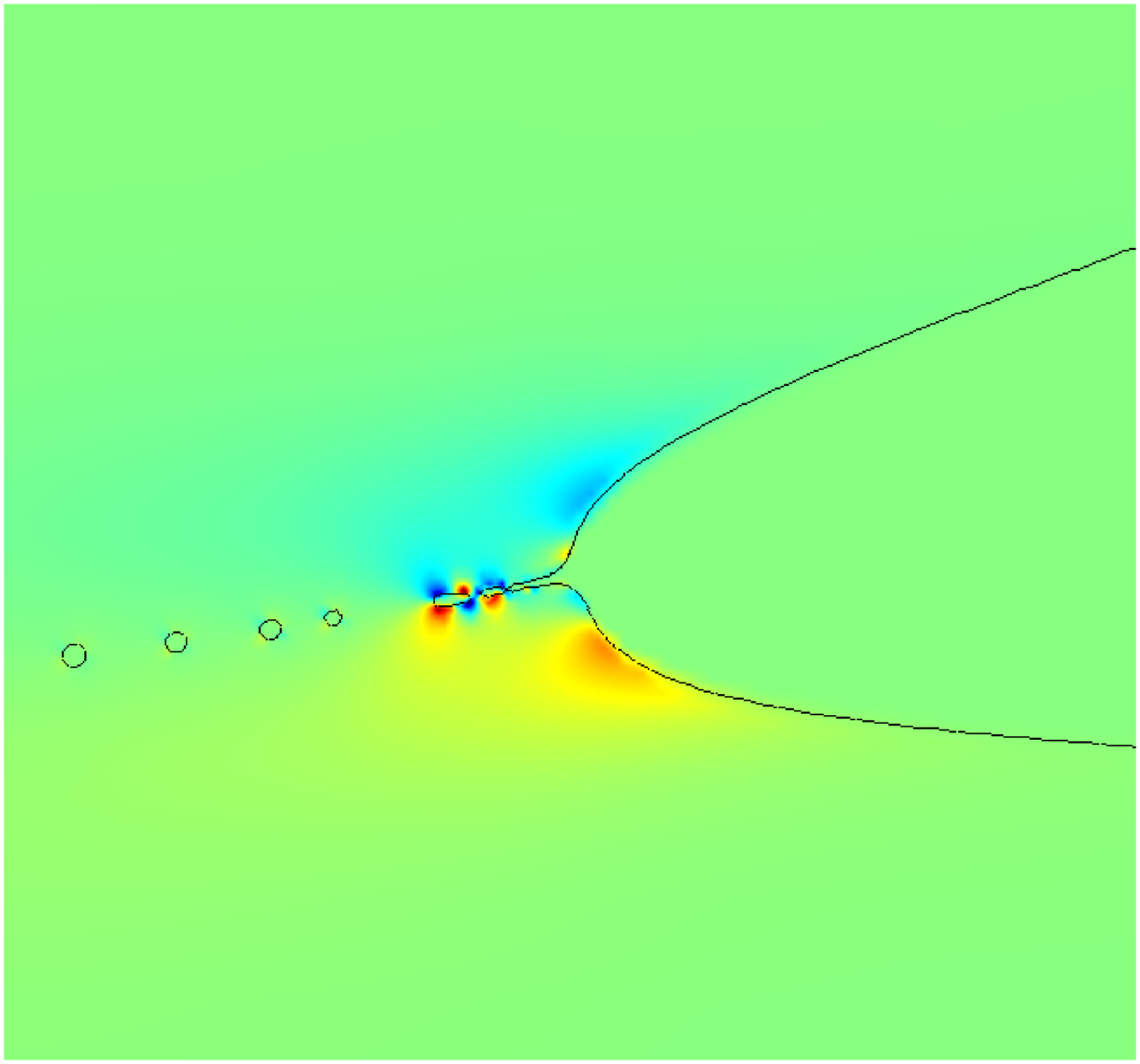}b)\includegraphics[width=6. cm]{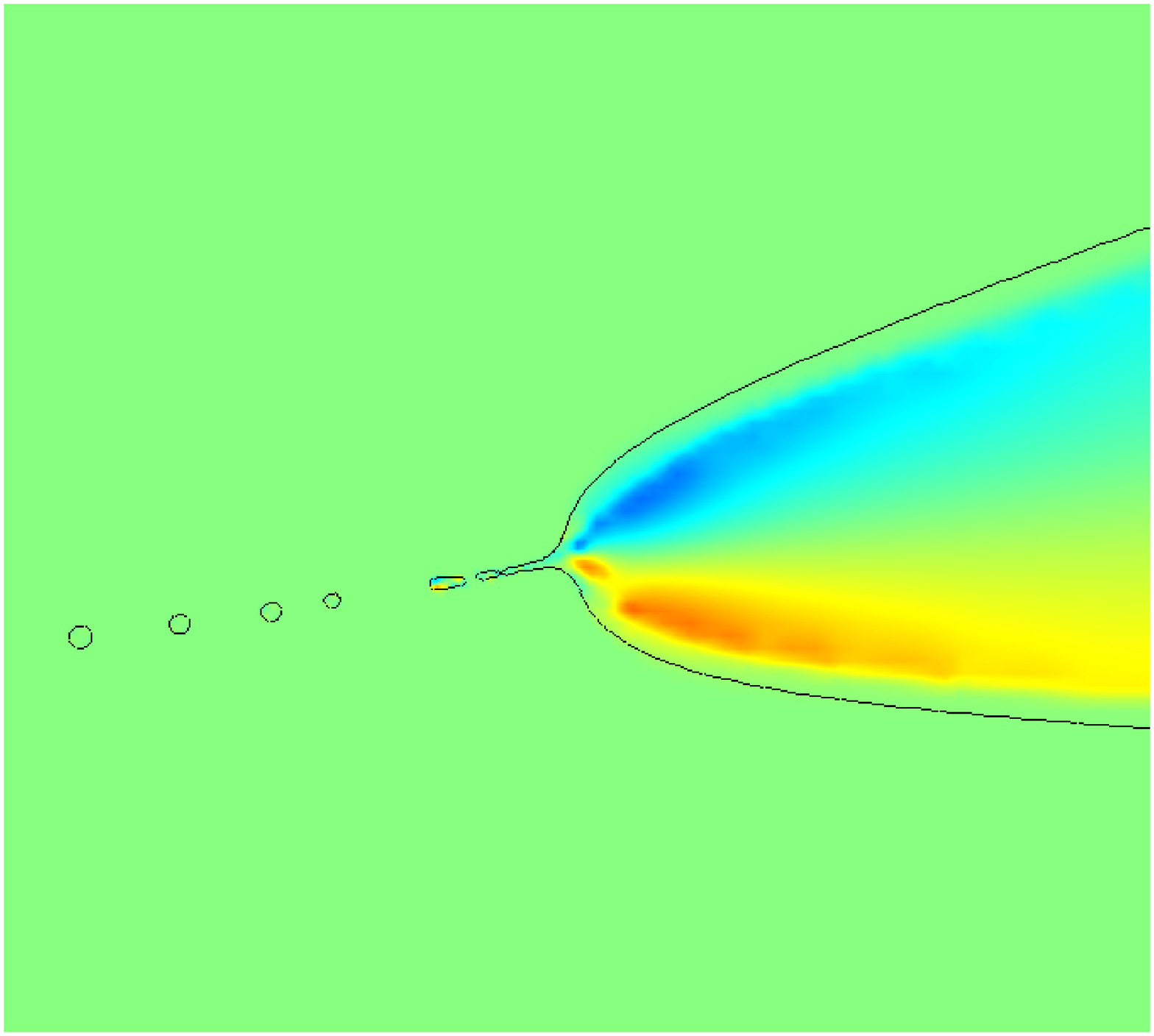}\\
\caption{The vorticity field corresponding to Figure \ref{zoom}d : a) in the liquid and b) in the gas.}
\label{zoomgridvort}
\end{figure}

In particular, small bubbles (which are actually toroidal because
of the axial symmetry) can be seen prior to the ejection of the thin
liquid sheet. In this case, the liquid of the droplet has already made contact with
the liquid layer before the jet formation.
Such small toroidal bubble entrapment might correspond to those
observed in experiments recently~\citep{thor12bubble}.
However it is also strongly controlled by the size of the grid, as 
reconnection in the VOF methods depends on the grid size. 
Here we note that with $\Delta x \sim 8\,10^{-8} m$ the grid size is one order of
magnitude away from the molecular length scales, so the VOF reconnection
although not physically realistic, may, in the future, approach 
the length scales at which molecular forces trigger reconnection in the real world. 
Finally, the mechanism of jet formation can be observed in Figure~\ref{zoomgridvort}, where the
vorticity field both in the liquid and gas phase is shown prior to the ejection corresponding to the zoom of Figure~\ref{zoom} d). It exhibits a vortex dipole at the origin of the jet, as already described in JZ03.

We can now characterize the splashing dynamics as the liquid and gas viscosities vary.

\subsection{Spreading radius}

\begin{figure}
\centering
\includegraphics[angle=-90,width=10.cm]{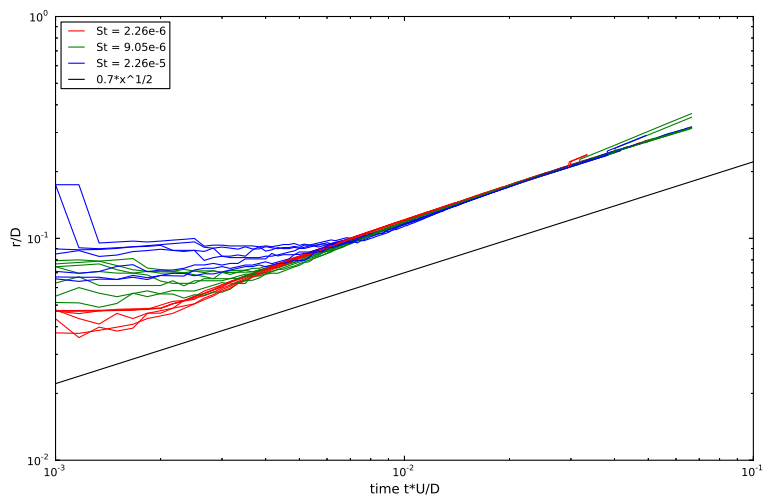}
\caption{The dimensionless spreading radius defined as the radius where the velocity is maximal in the liquid, as function of the dimensionless time $U_0t/D$, in a log-log plot for all the simulations performed in
this study. The straight line indicates the
slope $1/2$ corresponding to the geometrical law $r_g=\sqrt{DU_0t}$ shifted below for eye guiding.
As suggested by this geometrical relation, the different curves collapse all on a single one 
almost parallel to the expected law, showing that viscous, capillary and lubrication effects only
alter slightly this dynamics. Differences can however be seen at short 
times.}
\label{spreadrad}
\end{figure}

One of the crucial quantities involved in the scaling analysis is the geometrical radius $r_g(t)$ that acts as the
horizontal length scale. In order to verify that the horizontal scale behaves like $r_g(t)$, we investigate the evolution with time of the spreading radius, defined as the point in the liquid where the velocity is maximal. Figure~\ref{spreadrad} shows the evolution of the spreading radius with time for all the simulations performed. The square-root scaling ($r_g(t)=\sqrt{DU_0 t}$) is
observed over a large range of time with the same prefactor for all simulations.
Remarkably, the figure shows that this geometrical argument for the horizontal characteristic length is particularly robust and that the liquid properties (viscosities, densities) only influence the dynamics at short times. 

\subsection{Initial gas sheet formation}
We now study the formation of the gas sheet and its scaling. 
The transition from free fall to air cushioning can be seen on the time history of the heights of the film and the bubble
$z_-(r,t)$ and $z_+(r,t)$. 
Figure~\ref{z_t} shows both heights $z_-(0,t)$ and $z_+(0,t)$ on the axis as a 
function of time for  $\St=2.26 \, 10^{-5}$ and $\re = 1000$. This corresponds to 
a value of the air viscosity $18$ times its ordinary value, while the liquid is $8$ times more viscous than water.
It is seen that the heights behave linearly until some time near $t=0$ (approximately $t\sim -10^{-3}$. 
The linear behavior of $z_{+}(0,t)$ before impact is an indication that the bottom of the 
drop falls at the free fall velocity $U_0 + {\cal O}(g(t-t_0))$ (the gravity correction is even
smaller than $\Fr^{-1}$ due to the short time of observation)  
almost unperturbed from its initial value. At $t=0$ on the other
hand the cushioning dynamics have fully set in. 
After time 0, $z_+(0,t) \sim z_-(0,t) \sim -U_0t/2$ and 
$z_+(0,t) - z_-(0,t)$ remains approximately constant. This half velocity linear decrease of both $z_+(0,t)$ and $z_-(0,t)$ can be understood simply by momentum conservation 
as already suggested by~\cite{Tran2013}.
The scale $h_b$ of the gas layer may thus conveniently be defined as
$h_b = z_+(0,0) - z_-(0,0)$.

\begin{figure}
\centering
\includegraphics[width=10.cm]{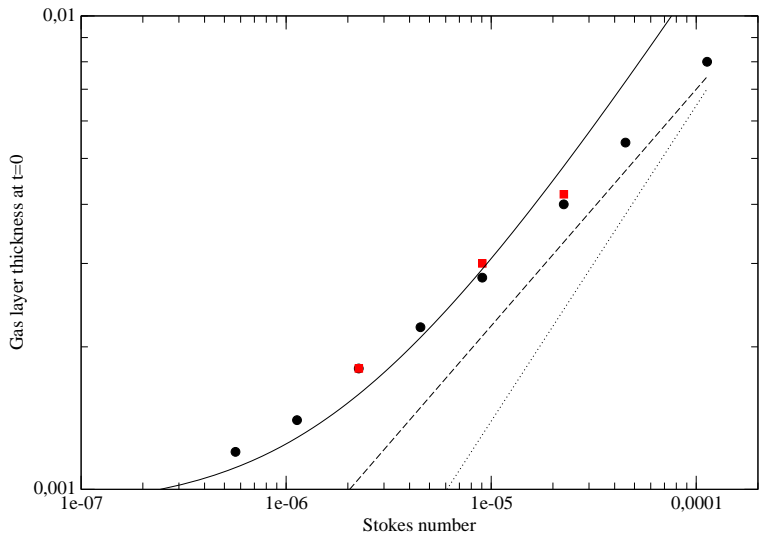}
\caption{Height scale $h_b/D$ of the gas layer as a function of the Stokes number for two values of the 
Reynolds number $\re=800$ (red square) and $\re=2000$ (black circle). The fit using equation (\ref{hbtheory}) is shown (solid line) as well as
the two scaling laws $St^{2/3}$ (dotted line) and $St^{1/2}$ (dashed line).}
\label{h_St}
\end{figure}

To determine the scaling of $h_b$ two series of simulations have been performed 
at $\re=2000$ and $\re=800$ for variable Stokes number $\St$. 
Together with the numbers in Table~\ref{allparams} these completely define the simulation parameters. 
The dimensionless height $\hat h = h_b/D$ is plotted on Figure~\ref{h_St}
together with relation (\ref{hbtheory}). 
The unknown turbulent friction coefficient $C_1$ has been fitted by trial and error to 
$C_1 = 0.75 \pm 0.1$. 

\begin{figure}
\centering
\includegraphics[width=10.cm]{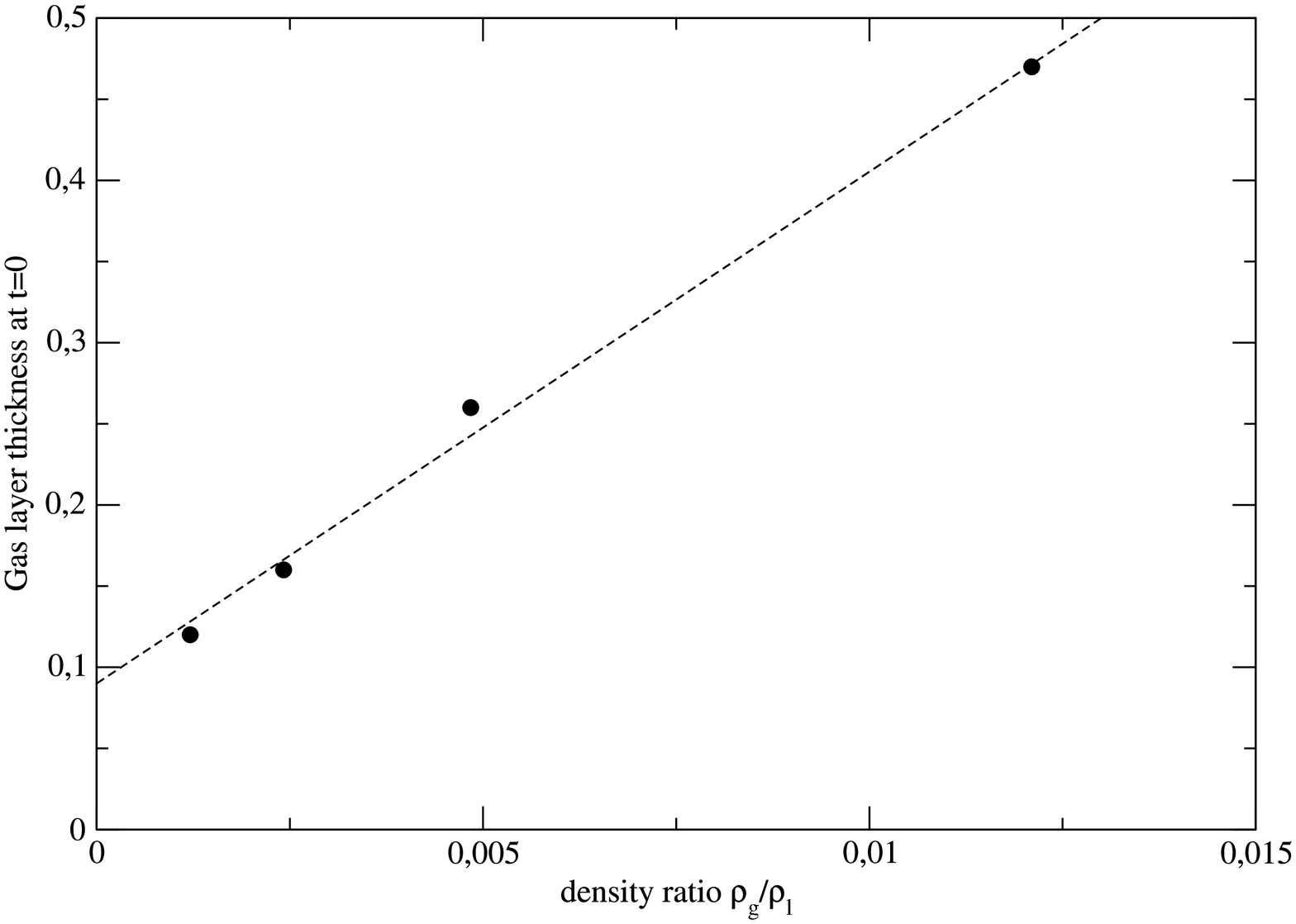}
\caption{Height scale $h_b/D$ of the gas layer as a function of $\rho_g/\rho_l$ for $\St = 5.67 \, 10^{-7}$
and $\re=2000$ and all other parameters as in Table~\ref{allparams}.}
\label{varrho}
\end{figure}

While the numerical data points are not exactly on top of the fit
the hypothesis of a transition from a $\hat{h}_b \rightarrow {\hat h}_{min}$ limit at small $\St$
to a $\hat{h}_b \sim \St^{n}$ behavior at larger (but still small $\St$) is compatible
with the data, with $n$ in some range around $2/3$. However, it is worth to remark that the rightmost part of the 
graph is closer to a $1/2$ power law behavior, suggesting that the alternative scenario proposed by~\cite{klaseboer2014universal} might be valid here. In fact, as the analysis on the pressure
will demonstrate it below, this alternative scaling is not valid here and the $St^{1/2}$ has to be seen 
as a best fit scaling in an intermediate regime~(\cite{taiwan15}).
In order to test the scaling of $\hat{h}_b$ at very low $\St$, when the effect of $\rho_g/\rho_l$ is most
marked, we perform a series of simulations at the smallest value of $\St$ in Figure~\ref{h_St}
and variable $\rho_g/\rho_l$, keeping all other numbers constant. The results are plotted on 
Figure~\ref{varrho}. We observe a linear increase of 
$\hat{h}_b$ with $\rho_g/\rho_l$, in agreement with the linear relation (\ref{hcr}), with a constant in the limit $\rho_g/\rho_l \rightarrow 0$ that depends on the Stokes number. 

\subsection{Impact pressure}

\begin{figure}
\centering
\includegraphics[width=\textwidth]{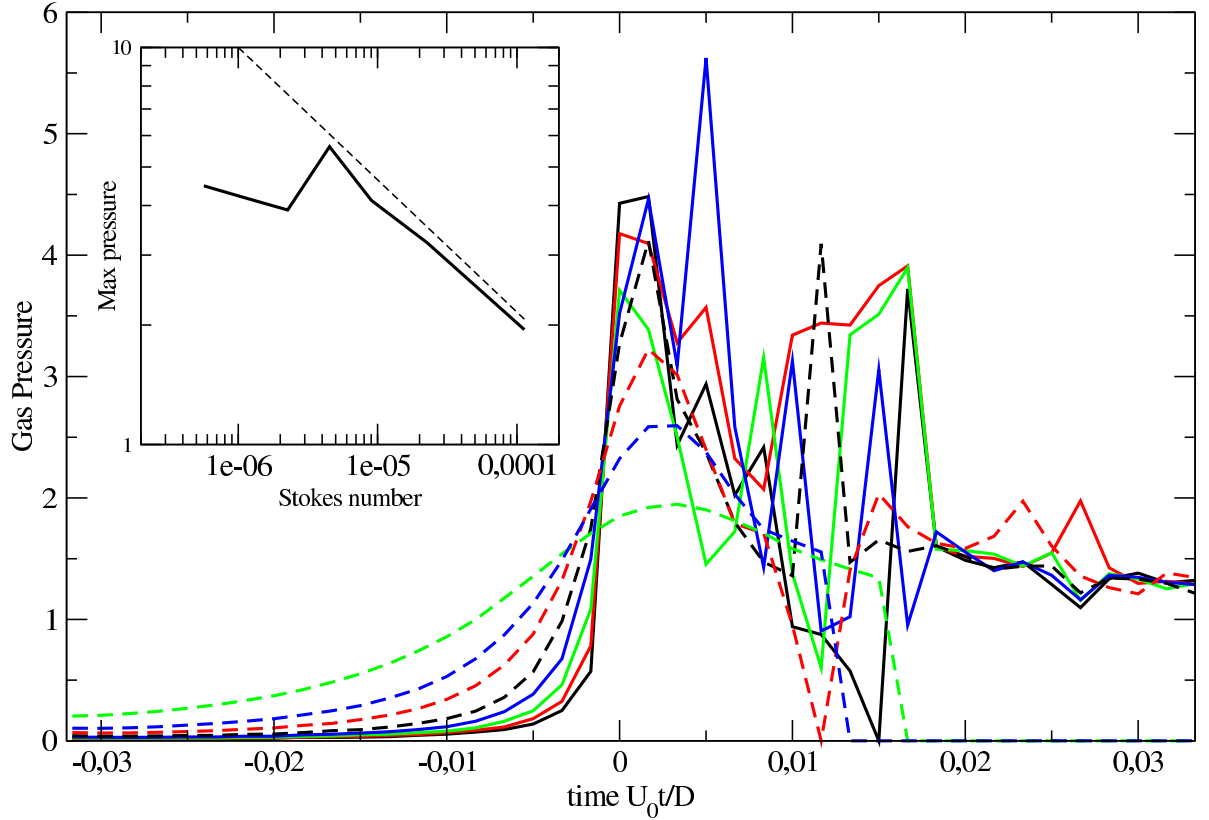}
\caption{The pressure in the gas layer in the cushioning regime defined as the maximum gas pressure on the axis down the drop. The dimensionless pressure (using $\rho_l U_0^2$) is plotted as a function of the dimensionless time $U_0 t/D$ for a fixed 
$\re=2000$ and for
varying Stokes number $\St=5.66\cdot 10^{-7}$, $1.13\cdot 10^{-6}$, $2.26\cdot 10^{-6}$, $4.52\cdot 10^{-6}$, $9.05\cdot 10^{-6}$, $2.26\cdot 10^{-5}$, $4.52\cdot 10^{-5}$ and $1.13\cdot 10^{-4}$, from top to bottom. The insert shows the maximum pressure over time of these curves as function of the Stokes number. The dashed line draws the expected $St^{-1/3}$ scaling following the prediction (\ref{max_p_imp})}
\label{pressure-uncol}
\end{figure}

In order to investigate quantitatively the various mechanisms
involved in the impact dynamics and the jet foramtion, we follow the evolution in time of the maximum
pressure on the axis in the gas layer as shown on Figure \ref{pressure-uncol} for different Stokes number at constant $\re=2000$. 
We see a very large pressure peak compared to the Bernoulli 
pressure $\rho_l U_0^2$. The insert of figure \ref{pressure-uncol} shows
a good agreement with the scaling $\St^{-1/3}$ predicted in (\ref{max_p_imp}). This dependence of the
pressure field with the Stokes number is clearly in disagreement with the Bernoulli argument proposed in~\cite{klaseboer2014universal}. Together with the variations of $\hat{h}_b$ found compatible with a 
$\St^{2/3}$ behavior, it indicates that the lubrication is the dominant regime in the air cushioning, by opposition to the alternative scenario of~\cite{klaseboer2014universal}. We should also emphasize here that this high pressure in the gas layer can lead to the compression of the gas. Indeed, taking for
instance the typical values for water drop impact $\rho_l=1000 \, {\rm kg}\cdot {\rm m}^{-3}$ and 
$U_0 \sim 3 \, {\rm m}\cdot {\rm s}^{-3}$, we obtain a pressure un the air of the order of one half of the
atmospheric pressure.

\subsection{Ejecta sheet velocity}

\begin{figure}
\centering
a\includegraphics[width=12cm]{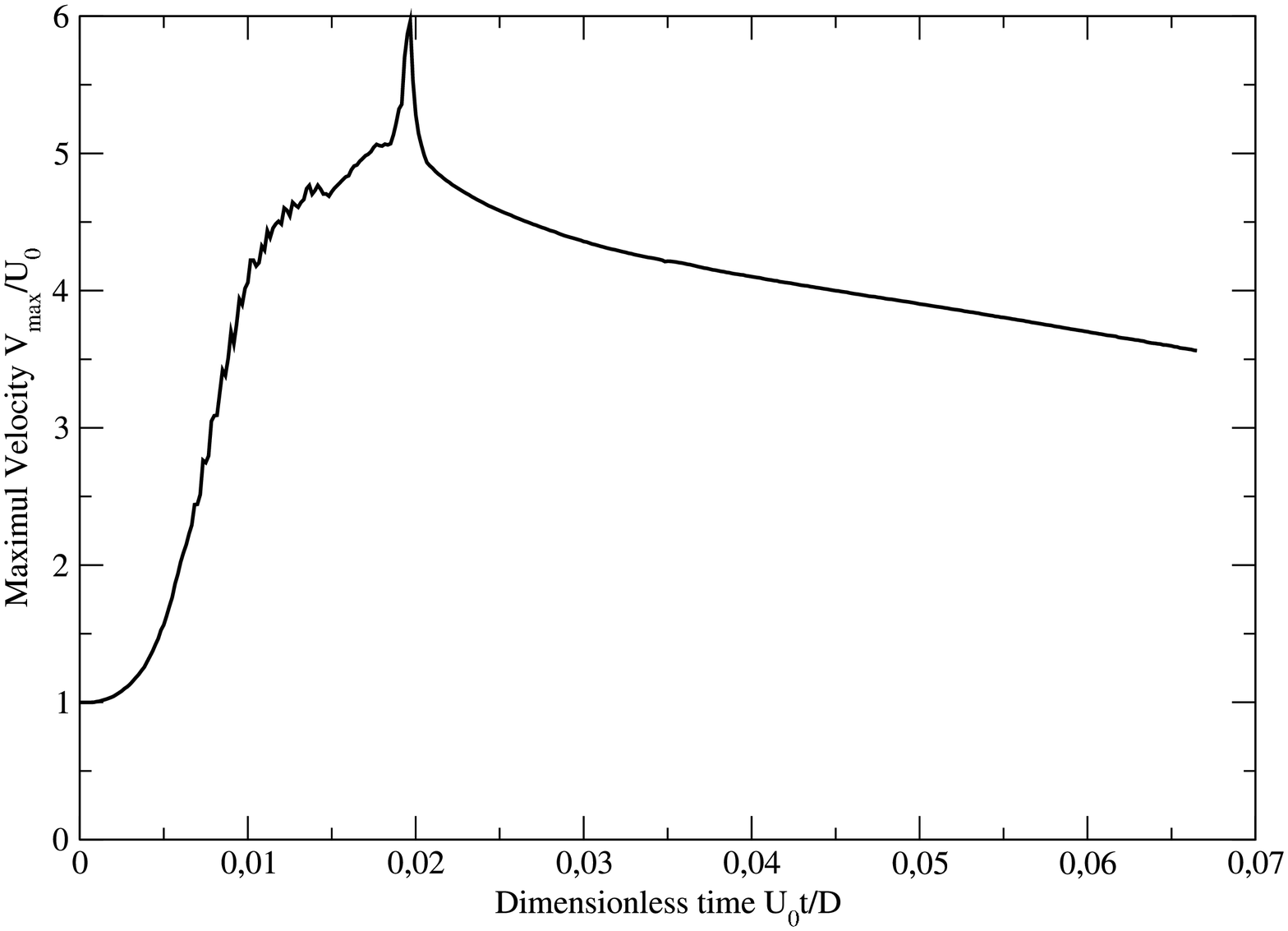}
\caption{The maximum velocity $V_{Max}/U_0$ as a function of the dimensionless time for  $\re=1000$ and $\St=9.05\, 10^{-6}$. 
Other parameters as in Table~\ref{allparams}.}
\label{velocity1}
\end{figure}

We study now the evolution of the
velocity maximum in the liquid.
This quantity is indeed an interesting proxy for several 
measurements. It is directly related to estimates of the velocity
of the jet itself. It is easier and less ambiguous to measure 
than the ejecta thickness, which varies widely at its base. 
Finally it has a telltale ``spike'' at the instant of jet ejection
marking $t_j$,
the time of emergence of the ejecta sheet.
On Figure~\ref{velocity1} we show the maximum 
dimensionless velocity as a function of the 
dimensionless time shifted to the beginning of
the simulation $U_0 t / D$ (remind that the origin of time $t=0$ corresponds to the time for which the geometrical falling sphere would interact with the liquid layer), for 
 $\re=1000$ and $\St=9.05\, 10^{-6}$, and other parameters in 
Table~\ref{allparams}. 
It is seen that the maximum velocity deviates from
the initial velocity $U_0$ around $-U_0t_b/D \simeq -0.03$, that is when the droplet approaches
the liquid film, after which it increases rapidly, then reaches a
maximum and decreases slowly. The maximum is often remarkably spiked. 

\begin{figure}
\centering
\includegraphics[width=6cm]{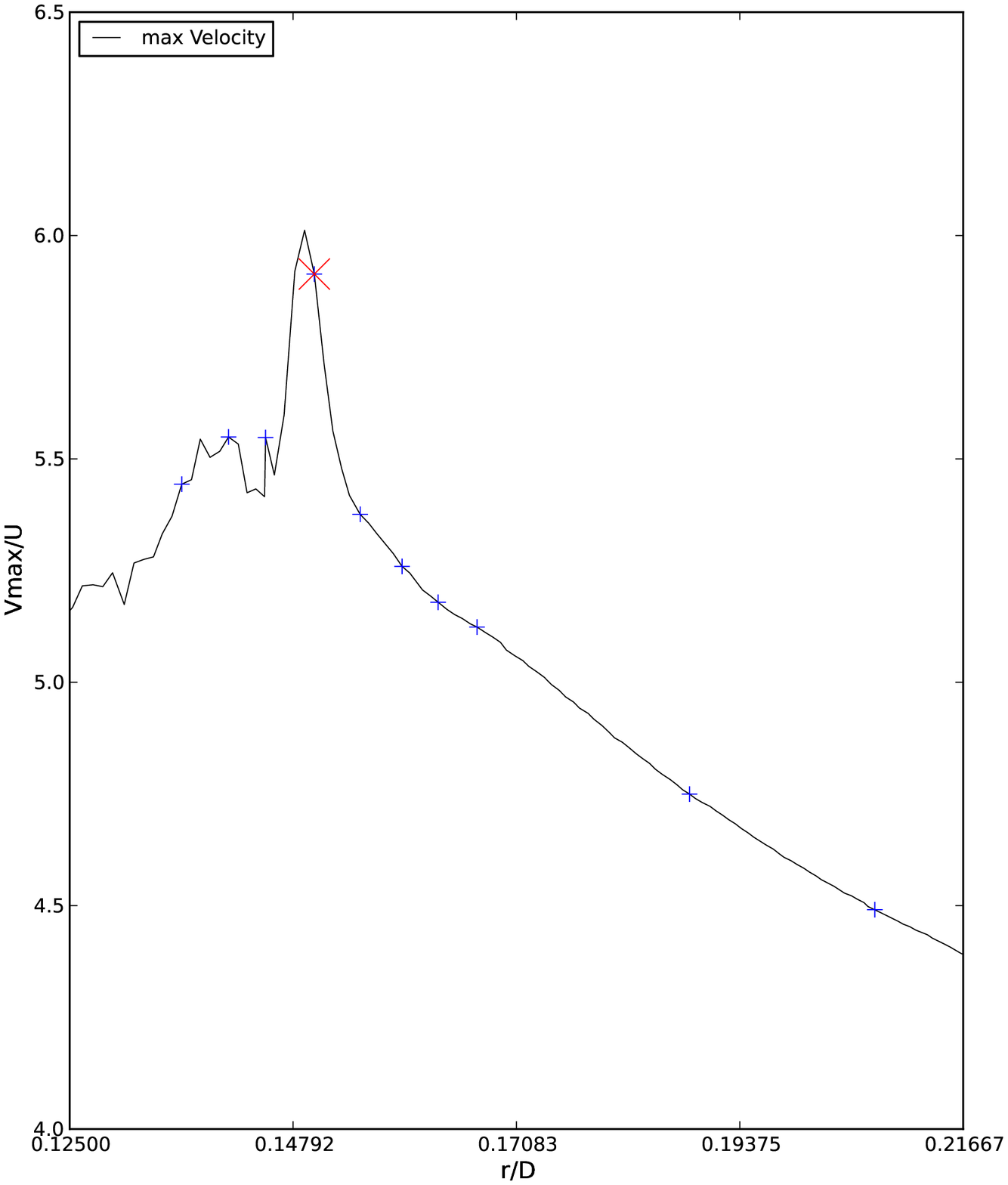}\includegraphics[width=6cm]{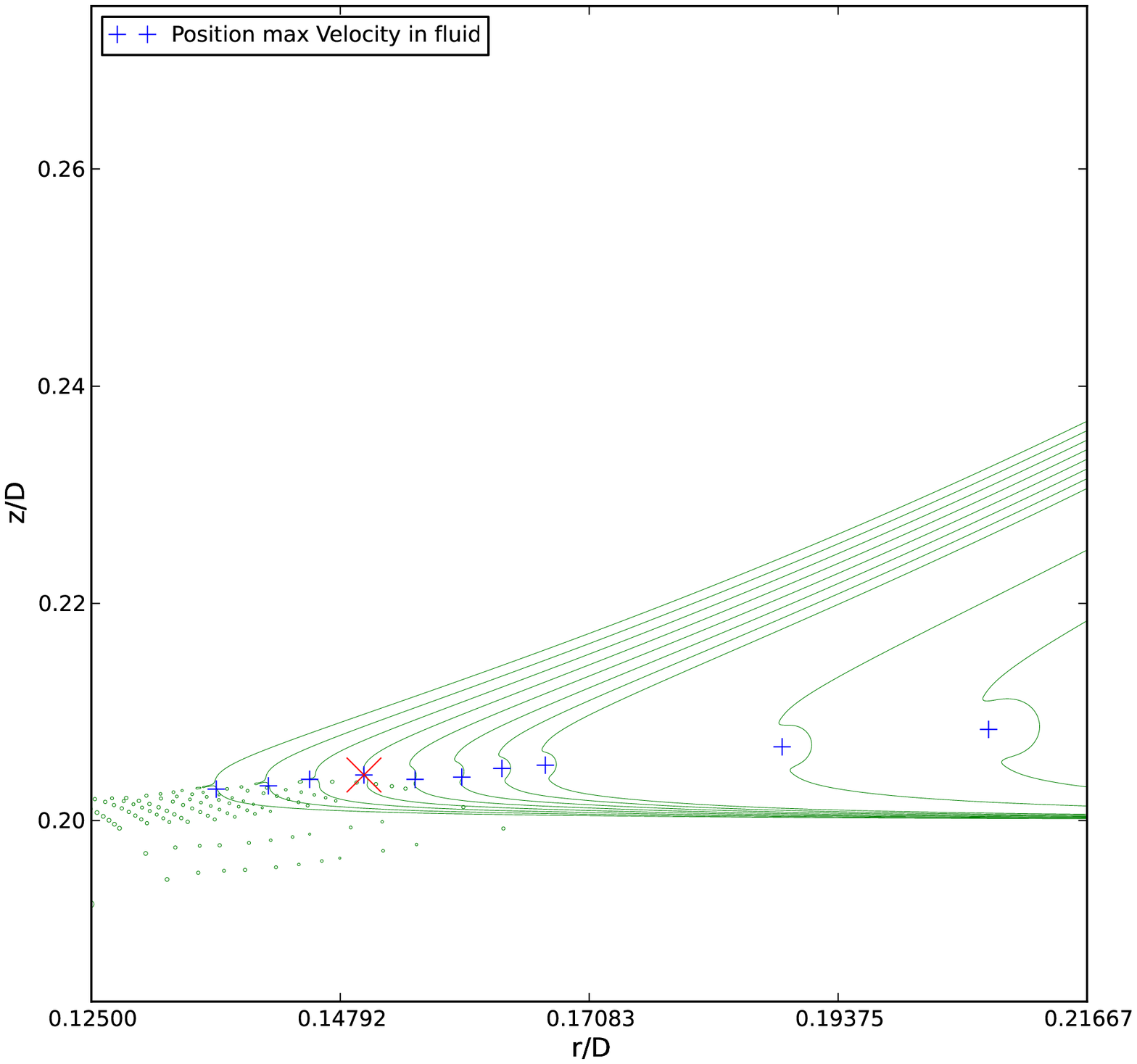}
\caption{Left: the maximum velocity as a function of the radius where it is located for different times (indicated by the blue signs).  Right: the interface profiles corresponding to the same time. The 
position at which the velocity is maximum
is again marked by blue $+$ marks. The location of the maximum velocity at time of reversal of the interface curvature, 
marking the beginning of jet ejection, is marked with a red $\times$ sign. 
}
\label{zoommaxvel}
\end{figure} 

Figure~\ref{zoommaxvel} shows the value
of the maximum velocity in the liquid  and the location at which it is reached 
as the time varies for the
same parameters.
It is clear in that case that the sharp peak corresponding to the maximum velocity
also corresponds to the time of reversal of the curvature of the interface, marking the 
beginning of the ejection of the jet. 
Detailed investigations show that in all cases investigated in this paper this spike
corresponds exactly to the time of formation of the ejecta sheet,
and that the maximum is located at its base. 

However, zooming on the base of the jet 
as done in Figure \ref{zoom}d shows that a set of
tiny bubbles is already formed, meaning that first contact between 
the drop and the sheet has already occurred {\em before} the jetting time in
Figure \ref{zoommaxvel}. In other words in that case contact happens
markedly before jet formation.

\begin{figure}
\centering
\includegraphics[width=12cm]{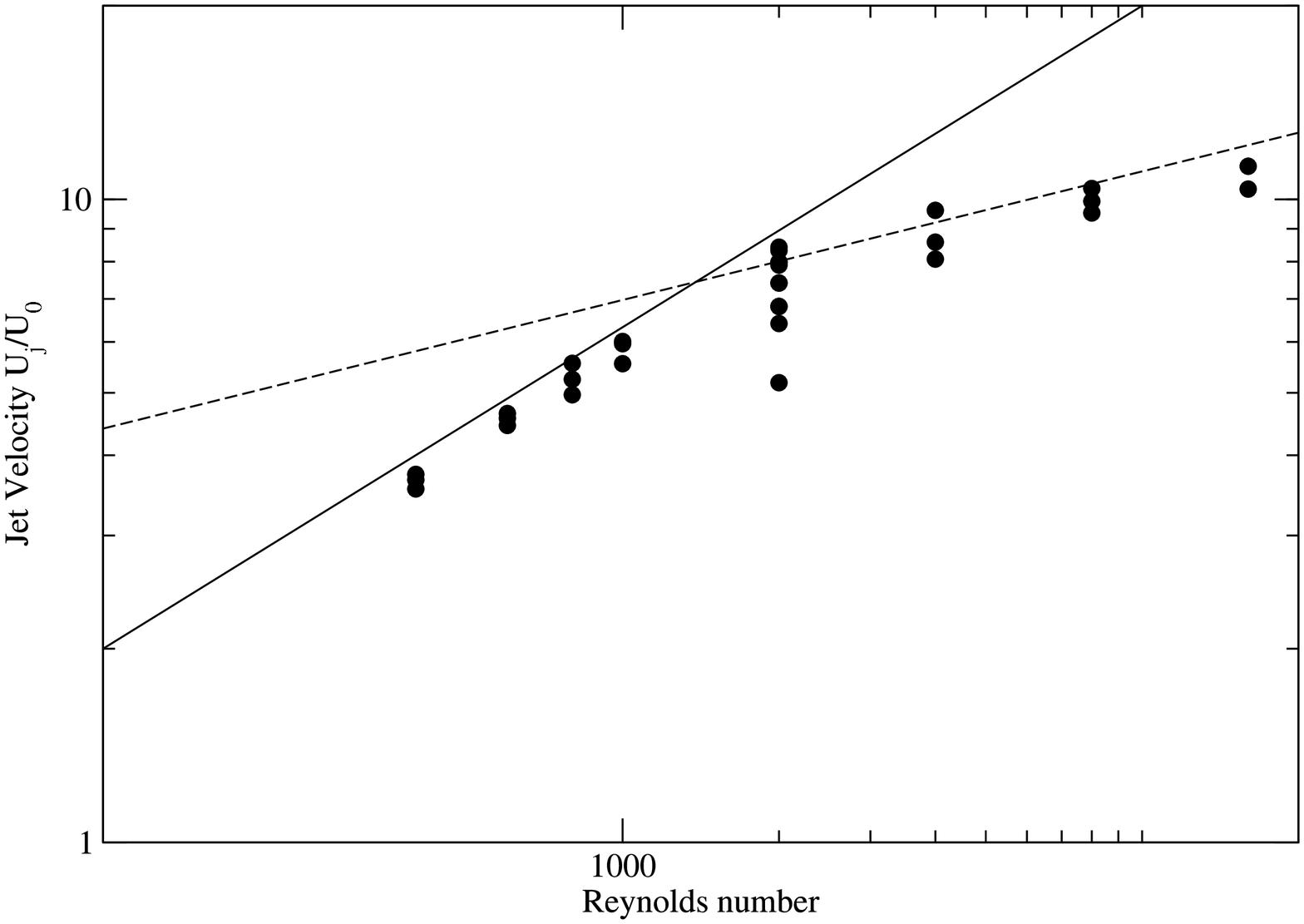} 
\caption{The jet velocities $U_j/U_0$ as function of the Reynolds number for all the simulations performed in this study, in a log-log plot.The predicted power law relation~\ref{jetvel} is plotted on showing only a reasonable agreement at Reynolds number lower than $1000$ (solid line). For higher Reynolds numbers another scaling appears, consistent with a $\re^{1/5}$ law (dashed line). For $\re=2000$, where many Stokes numbers have been considered, remark a slight dependence of the velocity with the Stokes number, following the higher the Stokes number, the lower is the jet velocity.}
\label{MaxVelo}
\end{figure}

Defining the speed of the jet as this spike velocity, we can investigate how the jet velocity depends 
on the Reynolds numbers for the different Stokes numbers simulated. This velocity is shown on 
Figure~\ref{MaxVelo} as function of the Reynolds number in order to check the validity of the scaling law~(\ref{jetvel}): $U_j \propto \sqrt{\re} U_0$. The results are somehow puzzling: indeed, as the 
predicted law exhibits a reasonably good agreement for "low" Reynolds number (below $1000$), 
important deviations appear at larger Reynolds where another scaling is apparently at play, consistent with a $\re^{n}$ fit with $n \sim 1/5$. However, it is interesting to notice that the jet velocity 
shows almost no dependence on the Stokes number below $\re\sim 1000$, as suggested by the viscous length theory of JZ03, while a small dependance can be identified in the higher Reynolds regime.

In order to better understand this discrepancy between the predicted law and the numerical results, 
the dimensionless time $\hat{t}_j=U_0 t_j/D$ of the jet formation needs to be investigated. Two scaling laws for this time are in 
competition: on the one hand the viscous length theory without accounting for the gas lubrication 
effect suggests that $\hat{t}_j \propto 1/\re$ (relation~\ref{tgcond}); on the other hand, the 
cushioning of the gas suggests that this effect is delayed by the time $\hat{t}_b \sim \St^{2/3}$ 
(relation~\ref{eqtb}).

\begin{figure}
\centering
\includegraphics[width=10cm]{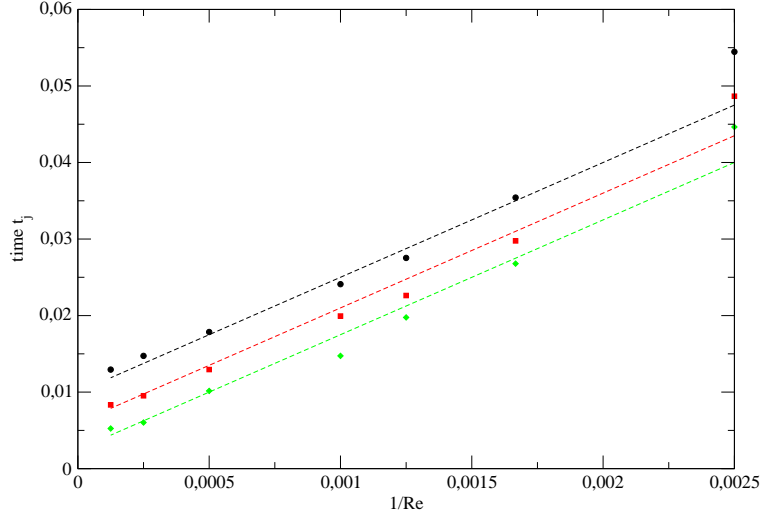}
\caption{${\hat t}_j=U_0 t_j/D$ as a function of the inverse of the Reynolds number ($1/\re$) for different Stokes number: black circle $\St=2.26\cdot 10^{-5}$, red square $\St=9.05\cdot 10^{-6}$ and green diamond 
$\St=2.26\cdot 10^{-6}$. Parallel dashed straight lines are drawn for each Stokes number for guiding the eyes.}
\label{tj-Re}
\end{figure}

\begin{figure}
\centering
\includegraphics[width=10cm]{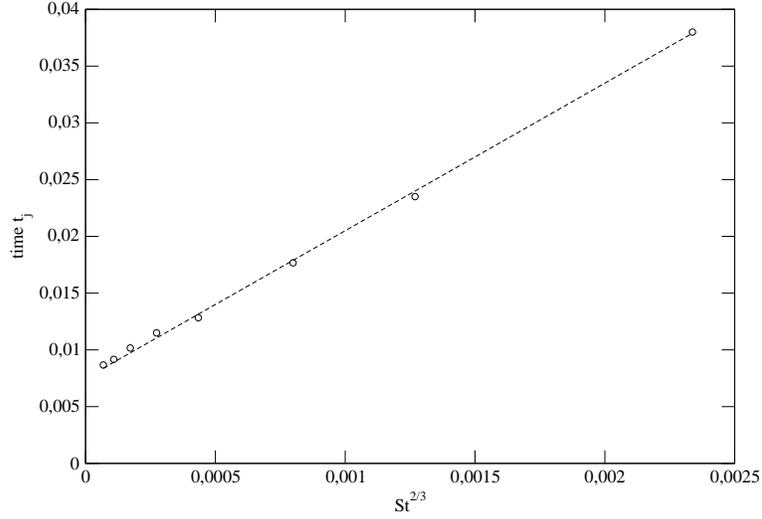}
\caption{${\hat t}_j$ as a function of the Stokes number to the power predicted by the theory (\ref{eqtb}) $\St^{2/3}$, for constant Reynolds number $Re=2000$. The dashed line indicates a linear relationship between these two quantities.}
\label{tj-St}
\end{figure}

First of all, Figure~\ref{tj-Re} exhibits parallel straight lines when plotting $\hat{t}_j$ as a function of $1/\re$ for three different Stokes numbers, indicating on the one hand that $\hat{t}_j$ evolves linearly with $1/\re$. On the
other hand, the straight lines are different for each Stokes number suggesting that a time delay 
depending on the Stokes number has to be considered. This can be seen on Figure~\ref{tj-St} where
$\hat{t}_j$ is plotted, for 
$\re=2000$, as a function of $\St^{2/3}$ as suggested by the theoretical law obtained for the bubble entrapment $\hat{t}_b$~(\ref{eqtb}). 

\begin{figure}
\centering
\includegraphics[width=10cm]{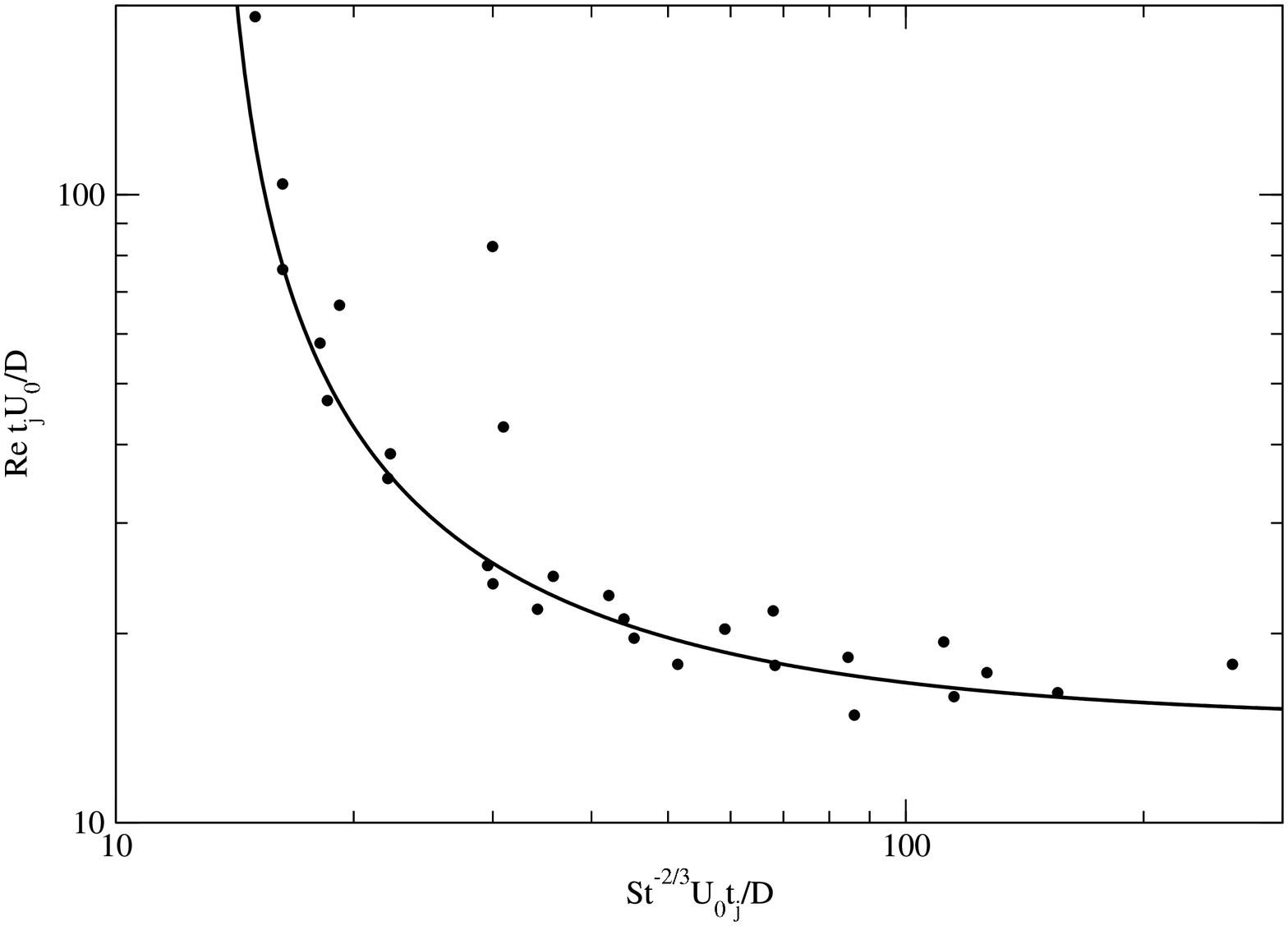}
\caption{The data (black circles) for the jet formation time plotted as the cloud of points $({\hat t}_j \re, {\hat t}_j \St^{-2/3})$ for all the simulations performed in this study. Formula~\ref{hypertj} is shown (solid curve) for $A_1=13.2$ and $B_1=14.5$.}
\label{alltjexp23}
\end{figure}

Again a nice straight line is observed, demonstrating eventually that the $\hat{t}_j$ obeys
the following relation:

\begin{equation} 
{\hat t}_j = A_1 \St^{2/3} + B_1 \re^{-1} \label{supertj}
\end{equation} 
which is equivalent by elementary algebra to 
\begin{equation}
{\hat t}_j\re = B_1 + \frac{A_1 B_1}{{\hat t}_j\St^{-2/3} - A_1} \label{hypertj}
\end{equation} 
with $A_1$ and $B_1$ fitting parameters. Figure \ref{alltjexp23} confirms this relation, by plotting
${\hat t}_j\re$ as a function of ${\hat t}_j\St^{-2/3}$, together with the latter relation~(\ref{hypertj}) for 
$A_1=13.2$ and $B_1=14.5$, showing a good collapse of the data on this master curve.

From Figure \ref{tj-Re} one can argue that the fitting constants $A_1$ and $B_1$ may eventually 
depend slightly on the Reynolds and Stokes numbers respectively. However, no clear and strong 
trend could be extracted from the data available so far and we will consider $A_1$ and $B_1$ as constant as a first approximation.
The above study suggests that the impact dynamics can be decomposed in two dynamical stages:
a first one dominated by the cushioning dynamics involving a $\St^{2/3}$ time scale dependence. 
Then, the ejection mechanism of the liquid sheet arises after a time delay proportional to $1/\re$. 
\begin{figure}
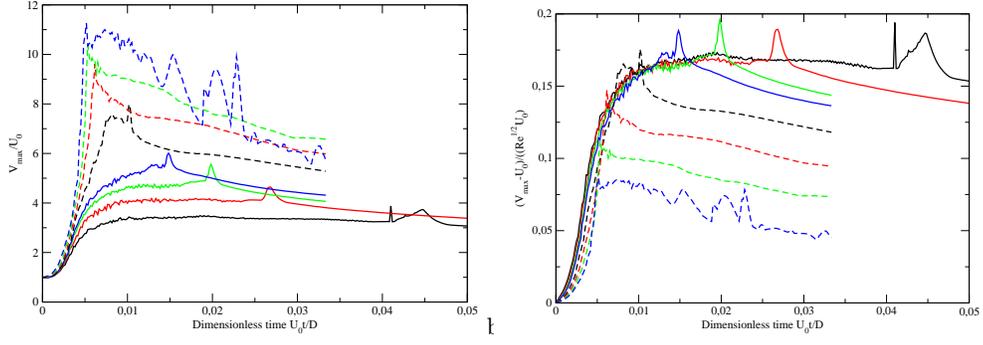

\centering
a)\includegraphics[width=6.25cm]{vitmaxSt226-6.eps} b)\includegraphics[width=6.25cm]{vitmrescaledSt226-6.eps}
\caption{a) Maximum velocity $V_{max}/U_0$ as function of time for $\St=2.26 \cdot 10^{-6}$ and for 
various Reynolds number ($\re=400$, $600$, $800$, $1000$, $2000$, $4000$, $8000$ and $16000$ 
for the curves from bottom to top). b) Same curves where the velocity has been rescaled by the 
theoretical prediction $\sqrt{\re}$. In this case the order of the curves for increasing Reynolds 
numbers are reversed, ranging from top to bottom. The initial velocity $U_0$ has been subtracted at 
short time for clarity.}
\label{rescaleRe}
\end{figure}

These different regimes can now be investigated through the evolution of the maximum velocity as 
function of time for all the parameters simulated here. Firstly, we show on figure \ref{rescaleRe} a), the
maximum dimensionless velocity $\hat{V}_{max}=V_{max}/U_0$ as function of the dimensionless time $\hat{t}=U_0t/D$ for different Reynolds number for a fixed Stokes number $\St=2.26 \cdot 10^{-6}$,
where it can be observed that the higher the Reynolds number, the higher is the velocity as 
expected by the JZ03 prediction (\ref{jetvel}). This is investigated in figure \ref{rescaleRe} b), where
these dimensionless velocities rescaled by the predicted scaling $\sqrt{\re}$ are plotted as function of time. If a reasonable collapse of the curve is obtained firstly at short times and for Reynolds numbers
below $1000$, the curves for higher Reynolds numbers deviates from the master curve starting at
the dimensionless time $\hat{t}_j$ of the jet formation. As expected however, the time $\hat{t}_j$ decreases as the Reynolds 
number increases, but for the high Reynolds numbers, the velocity peak appears during the 
velocity rise indicating that eventually the two mechanisms of air cushioning and jet formation interact.
Somehow, the air cushioning effect is interrupted by the ejection of the liquid sheet.
This could explain why the jet velocity at high Reynolds number does not follow the
prediction (\ref{jetvel}). On the other hand, for 
lower Reynolds number, one can see that the two mechanisms of air cushioning and the jet formation are well separated in time.

\begin{figure}
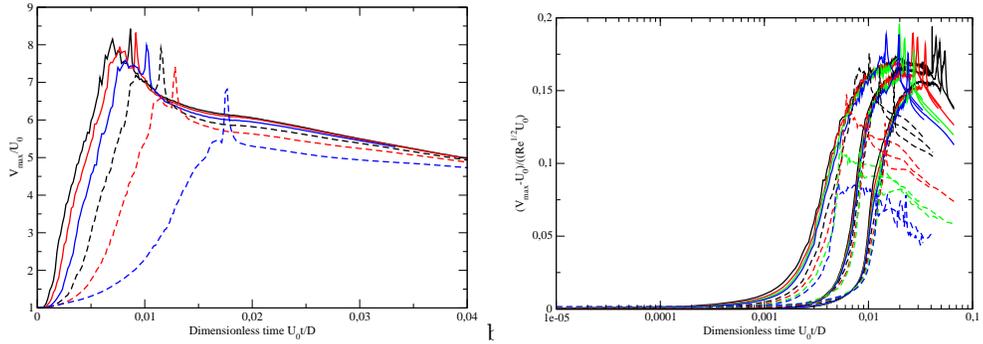

\centering
a)\includegraphics[width=6.25cm]{VmaxRe2kAllSt.eps} b)\includegraphics[width=6.25cm]{vitmrescaledRe3St.eps}
\caption{a) Maximum velocity $V_{max}/U_0$ as function of time for $\re=200$ and for various Stokes 
numbers ($\St=5.65\cdot 10^{-7}$, $1.13 \cdot 10^{-6}$, $2.26 \cdot 10^{-6}$, $4.52 \cdot 10^{-6}$,
$9.05 \cdot 10^{-6}$ and $2.26 \cdot 10^{-5}$ for 
the curves from left to right). b) For three Stokes numbers $2.26 \cdot 10^{-6}$, 
$9.05 \cdot 10^{-6}$ and $2.26 \cdot 10^{-5}$ the maximum velocity rescaled by $\sqrt{\re}$ are shown for different Reynolds numbers ranging from $400$ to $16000$, in a log-linear plot. Three sets of curves can be identified, one for each Stokes number, from left to right as the Stokes number increases. Here, the initial velocity $U_0$ has been subtracted at short time for clarity.}
\label{rescaleSt}
\end{figure}

The dependence of the dynamics on the Stokes numbers can be observed on figure \ref{rescaleSt} a) 
where the maximum velocity $V_{max}/U_0$ is shown at $\re=2000$ for different Stokes numbers.
As expected by the lubrication theory, the higher the Stokes number the slower is the rise of the 
velocity. On the other hand, since the formation of the jet is delayed by this cushioning dynamics, we
observe that the velocity peak is also delayed and is slightly decreasing as the Stokes number increases. This is in 
contradiction with the JZ03 initial prediction (\ref{jetvel}) that was obtained neglecting the gas cushioning,
explaining why this relation is not verified. The velocity curves are rescaled on figure \ref{rescaleSt} b) by $\sqrt{\re}$ as suggested by the prediction (\ref{jetvel}) for three Stokes numbers and different 
Reynolds numbers (up to $8$ for a given Stokes number). The curves arrange in three sets (one for 
each Stokes number for all the Reynolds numbers) in time showing clearly that the Stokes number 
influences mostly the accelerating regime. The $\sqrt{\re}$ predicted scaling for the jet velocity 
is seen through the maximum of these curves that are very close one from each other.
However, as observed in the jet velocity curve shown in figure \ref{MaxVelo} and on figures {rescaleRe}, many 
curves do not reach this maximum because of the cushioning dynamics.

\begin{figure}
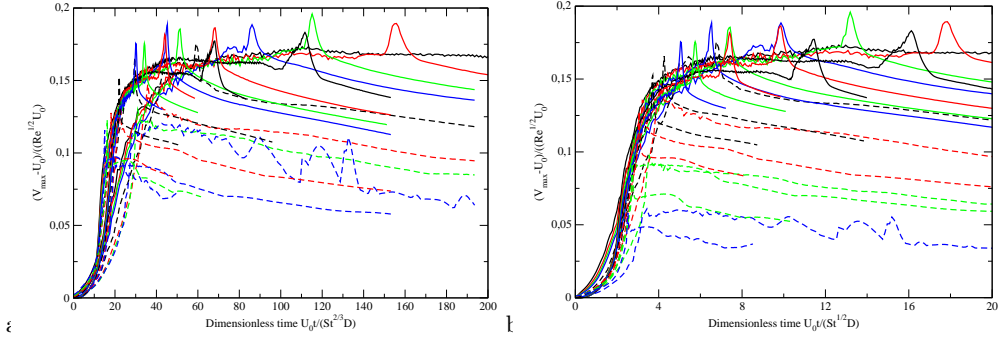

\centering
a)\includegraphics[width=6.25cm]{vitmallres23.eps} b)\includegraphics[width=6.25cm]{vitmallres12.eps}
%logt-filtr2_maxV_ejecta_scaling_allSt_withRe.pdf}
\caption{a) Rescaled velocities by $\sqrt{\re}$ for three Stokes number and up to eight Reynolds numbers shown as function of the rescaled time $U_0t/(D \St^{2/3})$ following the lubrication scaling $\hat{t}_b \sim \St^{2/3}$ \ref{eqtb}. b) Same curves but with the rescaled time involving a $\St^{1/2}$ scaling $U_0t/(D \St^{1/2})$
The initial velocity $U_0$ has been subtracted at short time for clarity.}
\label{rescaletSt}
\end{figure}

\begin{figure}
\centering
\includegraphics[width=12cm]{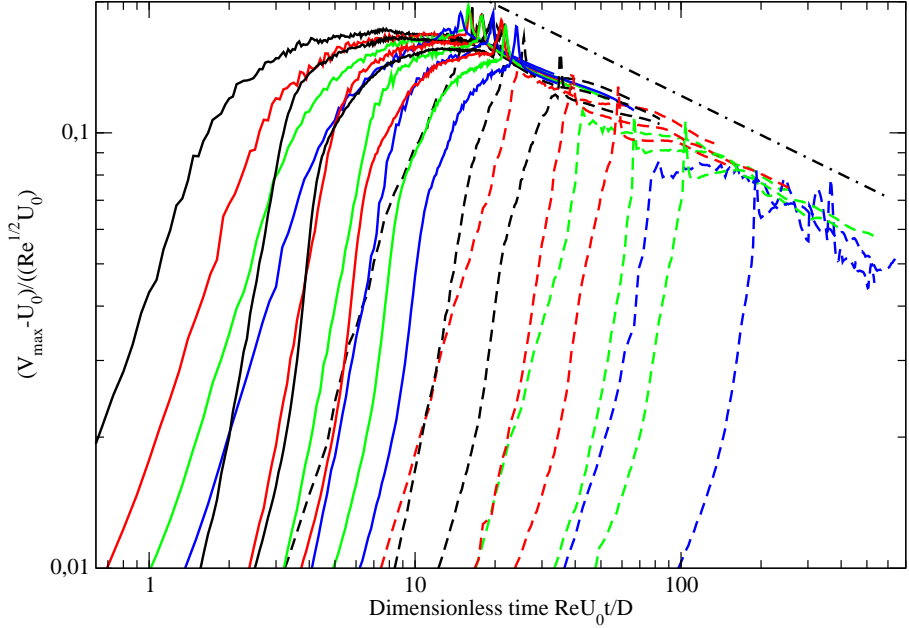} 
\caption{The rescaled maximum velocity $V_{max}/(\sqrt{\re} U_0$ shown as a function of the rescaled time $\re U_0 t/ D$ based on the geometrical timescale for jet ejection $t_g$ (see relation \ref{tgcond}) in a log-log plot. The dotted-dashed line indicates the power law scaling $x^{-3/10}$.
The initial velocity $U_0$ has been subtracted at short time for clarity.}
\label{rescaletRe}
\end{figure}

The curves for the three Stokes numbers on Figure~\ref{rescaleSt} 
result in three different curves which are roughly translated on the logarithmic time axis, suggesting a scaling dependence on the Stokes number as
proposed by the relation (\ref{jetvr}). Using exactly this scaling to rescale the time as 
$U_0t/(D \St^{2/3})$, figure \ref{rescaletSt} a) shows the maximum velocity for more than twenty 
different Stokes and Reynolds numbers. The collapse of all these curves is
 reasonably good, although a better collapse is obtained when rescaling the time by $\St^{1/2}$ 
 as shown on figure \ref{rescaletSt} b). This better collapse is somehow reminiscent of the $\St^{1/2}$ best fit 
 obtained for $h_b$ in figure \ref{h_St}. Therefore, the best collapse is obtained using a 
intermediate regime scaling where the time is best fit by $\St^{1/2}$. These results suggest
that in this first dynamical stage, where the air cushioning is dominant,
the velocity obeys the following relation:

\begin{equation}
V_{max} \sim \sqrt{\re} U_0 f_{c}[(U_0 t /D) \St^{-n}] 
\label{jetvr}
\end{equation}
valid {\it a priori before} the jet formation, where $n$ can be taken between $1/2$ and $2/3$. $f_{c}$ is the universal function that describes this
cushioning regime.
Immediately after the jet formation, and in a kind of plateau region, the velocity
at the base of the jet does scale with $\re^{1/2}$ for intermediate Reynolds numbers as predicted by 
Equation (\ref{jetvel}), while it departs slightly from this scaling for high Reynolds numbers. Although this evolution is in agreement
with the initial theory of JZ03, it is important to notice that the velocity is not constant and is in fact decreasing with time after the jet formation time $t_j$. This second stage is determined
by the jet dynamics and is {\it a priori} not influenced by the surrounding gas. Recalling that the 
timescale for the jet formation in the absence of gas cushioning is $t_g \sim D/(\re U_0)$, it is
tempting to investigate the evolution with time of the maximum velocity as a function of the rescaled time $t/t_g=\re U_0 t/D$, following:
\begin{equation}
V_{max} \sim \sqrt{\re} U_0 f_{j}(\re U_0 t/D) ,
\label{jetf}
\end{equation}
where $f_j$ is the universal function describing this second dynamical stage.
 The rescaled velocities $V_{max}/\sqrt{\re} U_0$ are shown on figure \ref{rescaletRe} as function of the rescaled time $\re U_0 t/D$ for all the simulations performed
in this study. Remarkably a nice collapse of the curves is observed for the large time dynamics {\it i. e.} for the time after the jet formation ($t>t_j$). The dashed line in this log-log plot shows a good fit of the data in this regime,
indicating a $-3/10$ power law, so that the maximum velocity obeys eventually for $t>t_j$:

\begin{equation}
V_{max} \sim \sqrt{\re} U_0 f_{j}(\re U_0 t/D)  \sim \sqrt{\re} U_0 \left(\frac{\re U_0 t}{D} \right)^{-3/10}.
\label{jetf1}
\end{equation}

This behavior suggests an explanation for the selection of the maximum jet velocity, considering that this latter regime starts at the jetting time $t_j$.
Recall firstly the asymptotic scalings observed for the jet maximum velocity shown on 
figure~\ref{MaxVelo}, namely $V_{max} \sim \sqrt{\re} U_0$ for $\re \le 1000$ and 
$V_{max} \sim \re^{1/5} U_0$ otherwise. Interestingly, these two asymptotics are consistent 
with the function $f_j$ (\ref{jetf1}) when considering the two asymptotics for the jet formation time $t_j$. Indeed, for $\re \le 1000$ we have $t_j \sim D/(\re U_0)$ so that, using \ref{jetf1}, the first scaling is$V_{max} \sim \sqrt{\re} U_0$. 
On the other hand, in the other regime, $\re \ge 1000$, for which $t_j \sim \St^{2/3}D/U_0$ we obtain: 
$$ V_{max} \sim \sqrt{\re} U_0 (\re \St^{2/3})^{-3/10} \sim \re^{1/5} \St^{-1/5},$$
giving remarkably the velocity scaling for large Reynolds numbers. Note also that it predicts a 
power law dependence on the Stokes number that we have not tested so far, although it can be qualitatively seen on figure~\ref{MaxVelo}.
Finally, using the formula (\ref{hypertj}) for $\hat{t}_j$:

$$ {\hat t}_j = A_1 \St^{2/3} + B_1 \re^{-1},$$

we obtain an effective formula for the maximum jet velocity:

\begin{equation}
V_{max} = C_1 \frac{\re^{1/5}U_0}{ (A_1 \St^{2/3}+\frac{B_1}{\re})^{3/10}} 
\label{eqfinal}
\end{equation}

\begin{figure}
\centering
\includegraphics[width=12cm]{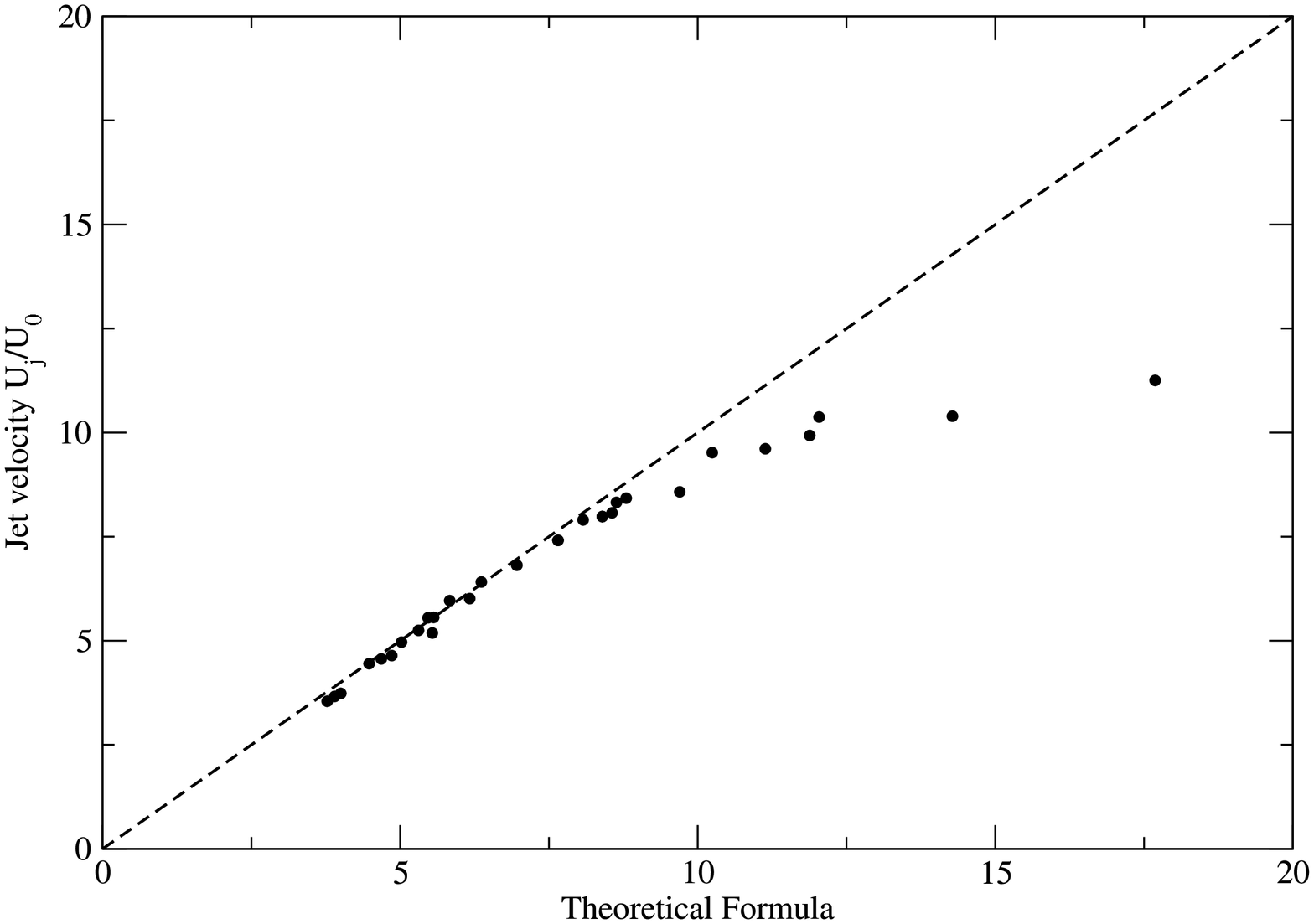} 
\caption{Black circles: the numerically measured jet 
velocity $U_j/U_0$ shown as function of the effective jet velocity eq. \ref{eqfinal} for all the cases studied here. The best fit is obtained for $C_1=0.45$ while $A_1=13.2$ and $B_1=14.5$ have been computed before. The dashed line shows the theoretical line $y=x$.}
\label{vitjpred}
\end{figure}

This formula is tested on figure \ref{vitjpred} where the measured jet velocity for all the 
numerical simulations performed here is compared to eq. (\ref{eqfinal}), where the best fit is obtained for $C_1=0.46$. If the results are correctly distributed on the line $y=x$ for the jet velocities up to $20$, it departs slightly from it for the higher velocities that correspond to the highest
Reynolds number studied here.

\section{Discussion and Conclusion}

In our numerical experiments, we have focused on the jet velocity and the effect of the gas layer. 
The formation and the scaling of the gas layer were analyzed. At the same time that the
gas layer forms, a large peak of pressure is observed which scales as 
$P_{imp} \sim \rho_lU_0^2 \St^{-1/3}$ 
as predicted by the lubrication theory that we have outlined in Section \ref{theory}, leading eventually to a bubble entrapment. 
The formation of the gas layer itself shows interesting symmetry properties between the 
dynamics of the  droplet surface $z_+$  and the dynamics of the liquid layer $z_-$. 
The two are equally deformed at the time $t=0$ defined geometrically as described above. 
The time during which deformation sets up is very short compared to the characteristic time
$D/U_0$ and also to the time of free fall $h_0/U_0$. The corresponding 
layer thickness is consistent with a $D\St^{2/3}$ scaling as outlined by the lubrication theory, although an intermediate inertial regime is also observed. Moreover, Figure \ref{pressure-uncol} demonstrates 
that the pressure is determined by the lubrication scaling, following $P_{imp} \sim 3 \rho_l U_0^2 \St^{-1/3}$, confirming that the dominant mechanism for the air cushioning is determined by the 
lubrication regime.
The results for the jetting time $t_j$ exhibit an interplay between the gas cushioning (whose 
time scale is determined by $\St^{2/3}$) and the 
liquid viscous boundary layer mechanism described in JZ03 (with a time scale in $1/\re$).
However, it is interesting to remark that the air cushioning dynamics could almost equivalently be fit 
by a $\St^{1/2}D/U_0$ time scale (instead of $\St^{2/3}D/U_0$), indicating that the intermediate inertial
regime is also observed.
The dependence of the layer thickness shows in addition a linear relationship with the density ratio $\rho_g/\rho_l$ (Figure \ref{varrho}) as predicted by the theory in 
equation (\ref{hcr}).

\begin{figure}
\centering
a)\includegraphics[width=6. cm]{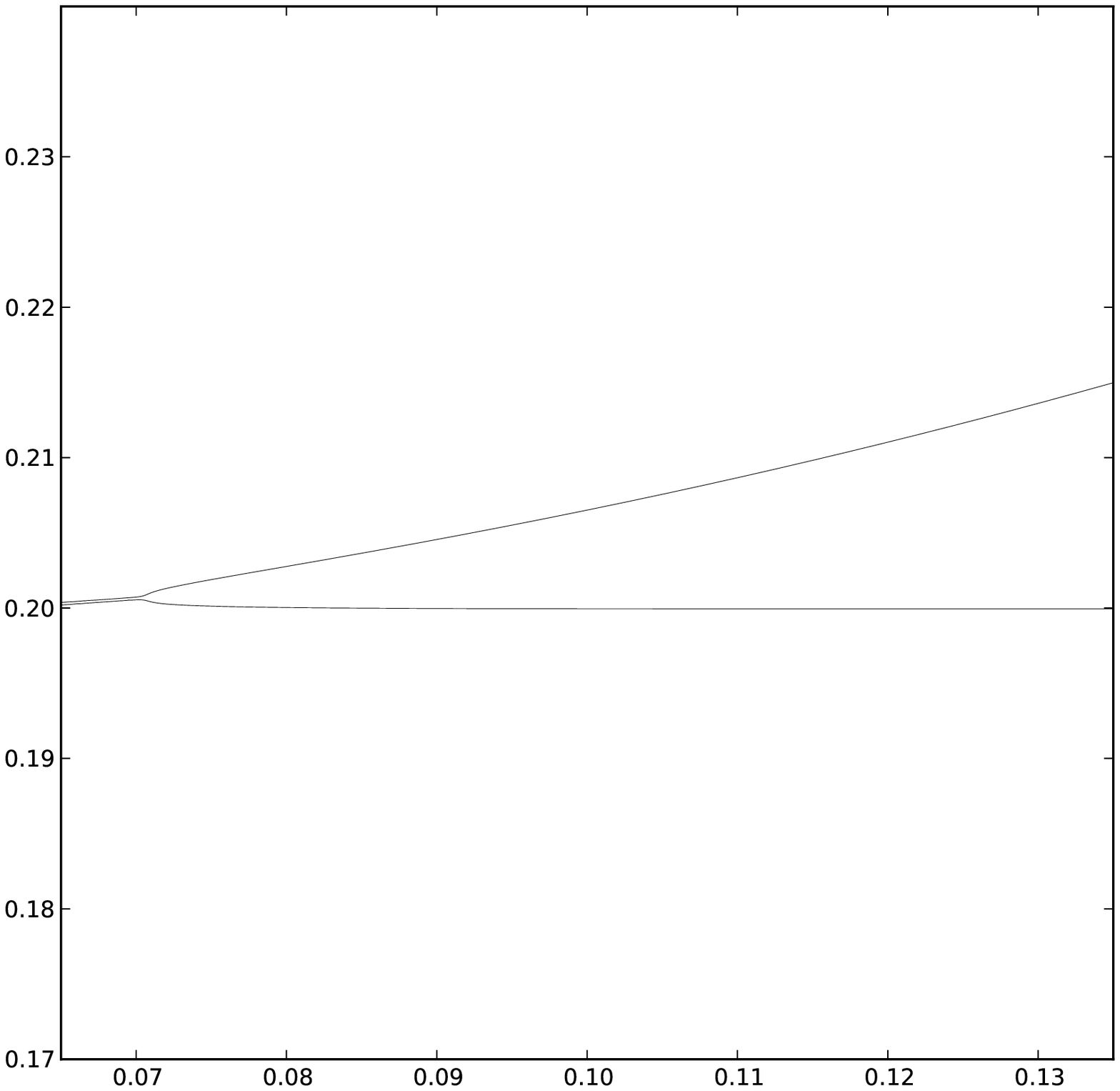}b)\includegraphics[width=6. cm]{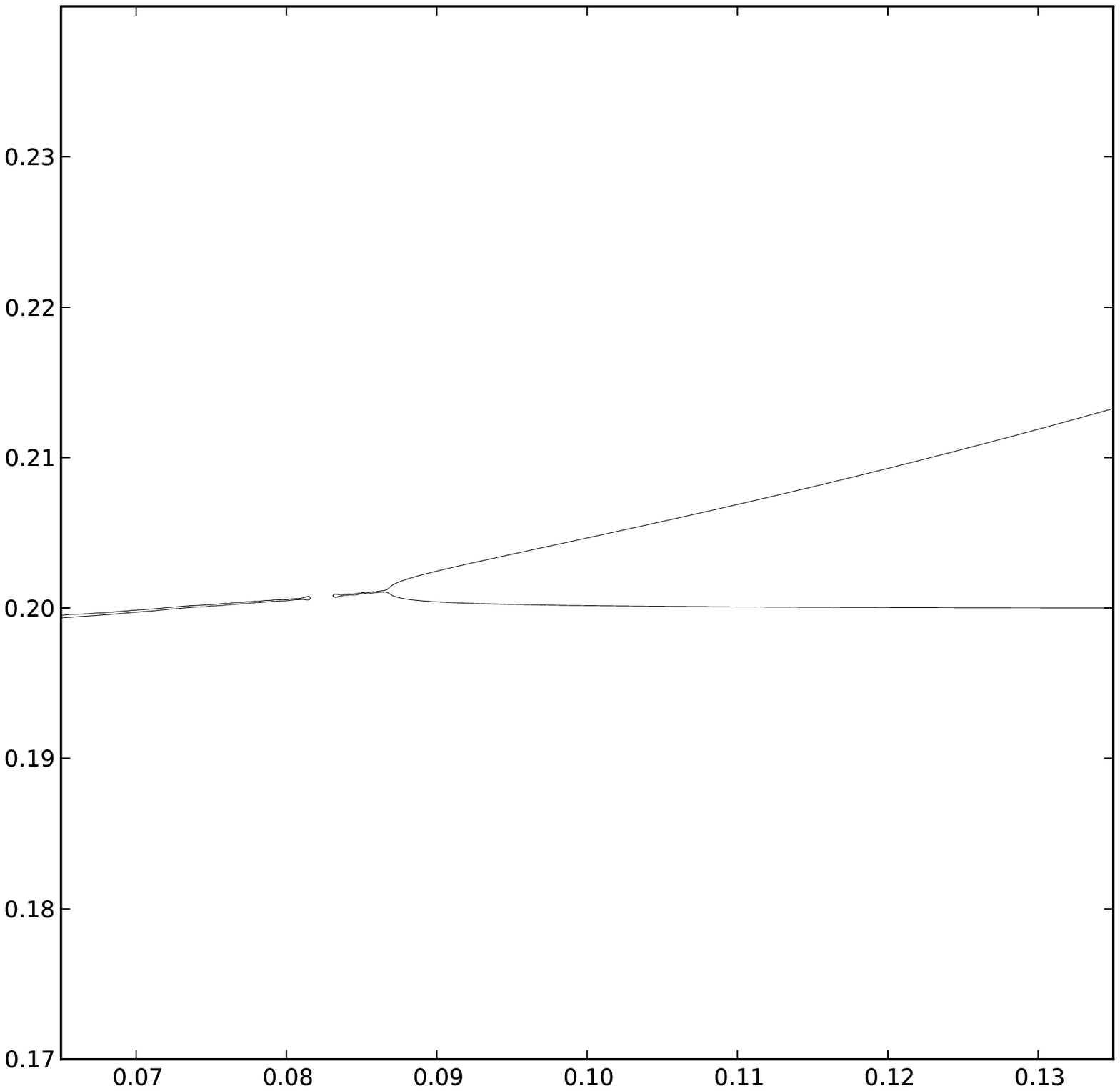}\\
c)\includegraphics[width=6. cm]{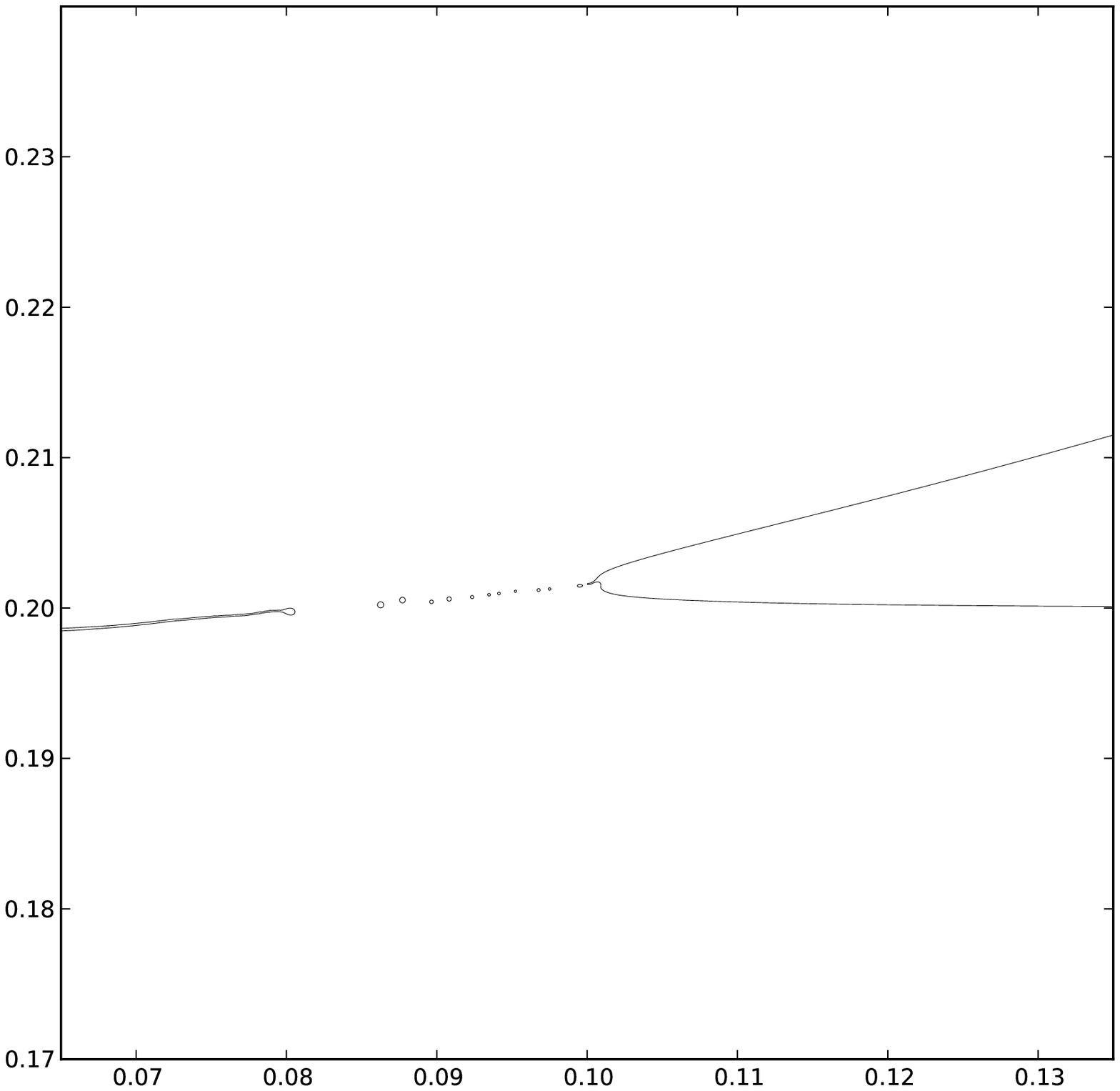}d)\includegraphics[width=6. cm]{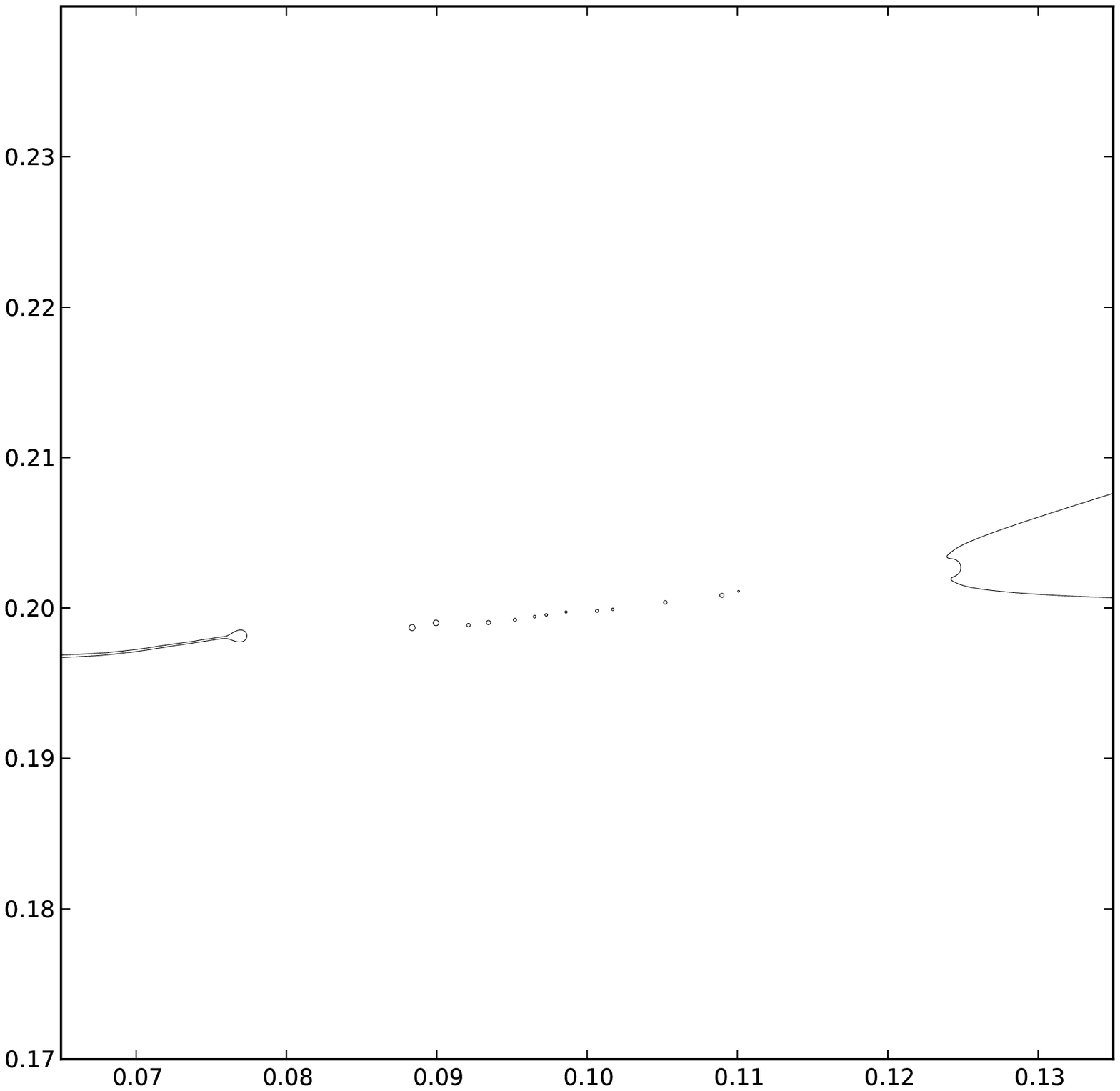}\\
\caption{Zoom on the connexion of the upper interface of the drop with the liquid layer for $\re=2000$ and $\St=5.66 \cdot 10^{-7}$, the other parameters being that of table \ref{allparams}. The white lines represent the interfaces and the colors scale in the liquid phase is realted to the velocity.The interfaces are shown on time $U_0t/D=3.67\cdot 10^{-3}$ a), $5.33\cdot 10^{-3}$ b), $7\cdot 10^{-3}$ c) and $1.03\cdot 10^{-2}$ d). The jet number $J=2.56 \cdot 10^{-3} \ll 1$ and we thus observe that after the first connexion of the two interface on figure a), small bubbles are created at the front of the connexion by the retraction of the thin
gas layer that is still between the falling drop and the liquid film, figure b) and c) until the jet emerges on figure d). On the other side of the connected region, the thin air film forming the entrapped large bubble is also retracting by capillarity.}
\label{figbulles}
\end{figure}

These results evidence a transition from a regime where the gas thickness and 
cushioning effect are insignificant on the jet dynamics to a regime where the air cushioning
controls the jet dynamics. From equation (\ref{hypertj}), we find that the air cushioning regime occurs for $\St^{2/3} \re \gg  1$ provided $\rho_g / \rho_l$ is small enough.
The corresponding number $\St^2 \re^3 = \re \mu^2_g/\mu^2_l$ varies only with $\re$ for a given
liquid-gas pair, so it fixes a limiting $U_0 D$ for which the regime changes 
for a given pair of fluids. For an  air-water system, 
taking $\mu_l/\mu_g \sim 50$ it gives a transition
at $\re \sim 3250$ or $U_0 = 3.25 m/s$ for a $D=1$ mmm droplet 
and $\mu_l = 10^{-3} {\rm Pa}\cdot {\rm s}$. The jet regime is thus characterized by a new
dimensionless number $J$ that is related to the ratio between the time for bubble entrapment $t_b$ to the geometrical time for the jet formation $t_h$: 
\begin{equation}
J=\St^2 \re^3=\left(\frac{t_b}{t_g}\right)^3
\label{Njet}
\end{equation}
We have observed in our simulations that in most cases the jet emergence is 
simultaneous with the connection of the two interfaces, that is the bubble becomes
trapped at the time of jet formation. In fact, the situation is more complex and can be analyzed using 
this jet number $J$ that quantifies the transition between a regime where the air cushioning is insignificant ($J \ll 1$) to a regime when it is dominant ($J \gg 1$).
When the air cushioning is insignificant, we find that the jet forms at the geometrical time $t_g \gg t_b$. with a jet velocity of the order of $\re^{1/2}$ as 
predicted earlier in JZ03. Moreover, we find that this velocity scale is present before jet formation 
indicating the existence of a large velocity in the droplet prior to the formation of the jet. 
This large velocity (asymptotically infinitely larger than $U_0$) is indicative of the focusing of the liquid 
velocity in a small region inside the droplet prior to the emergence of the jet.  
In the regime where the gas thickness and 
cushioning effect are insignificant at small $t_b/t_g$ the air layer has to close
before jet formation. A trace of that is seen in the presence of small bubbles on 
Figure~\ref{zoom}d before jet formation. This is even clearer on figure~\ref{figbulles} that shows the
details of the interface dynamics between the time of the first contact between the drop and the liquid film and the formation of the jet, for $\re=2000$ and $\St=5.66 \cdot 10^{-7}$, leading to $J=2.56 \cdot 10^{-3}$.

We observe that a delay exists between the first connexion of the interfaces (figure \ref{figbulles} a) 
and the jet formation (igure \ref{figbulles} d), during which small bubble are entrapped by the thin gas
film dynamics.

\begin{figure}
\centering
a)\includegraphics[width=6. cm]{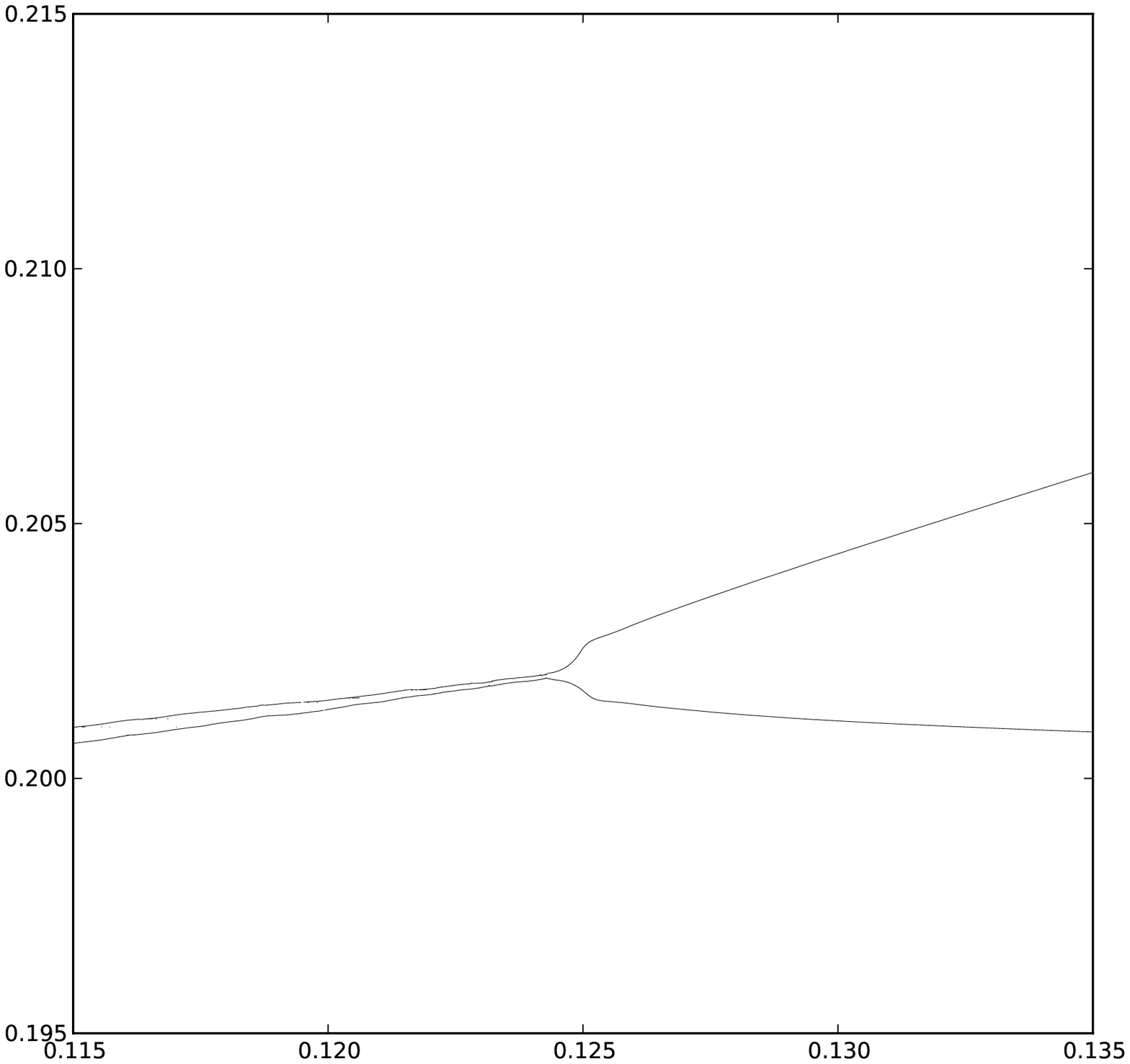}b)\includegraphics[width=6. cm]{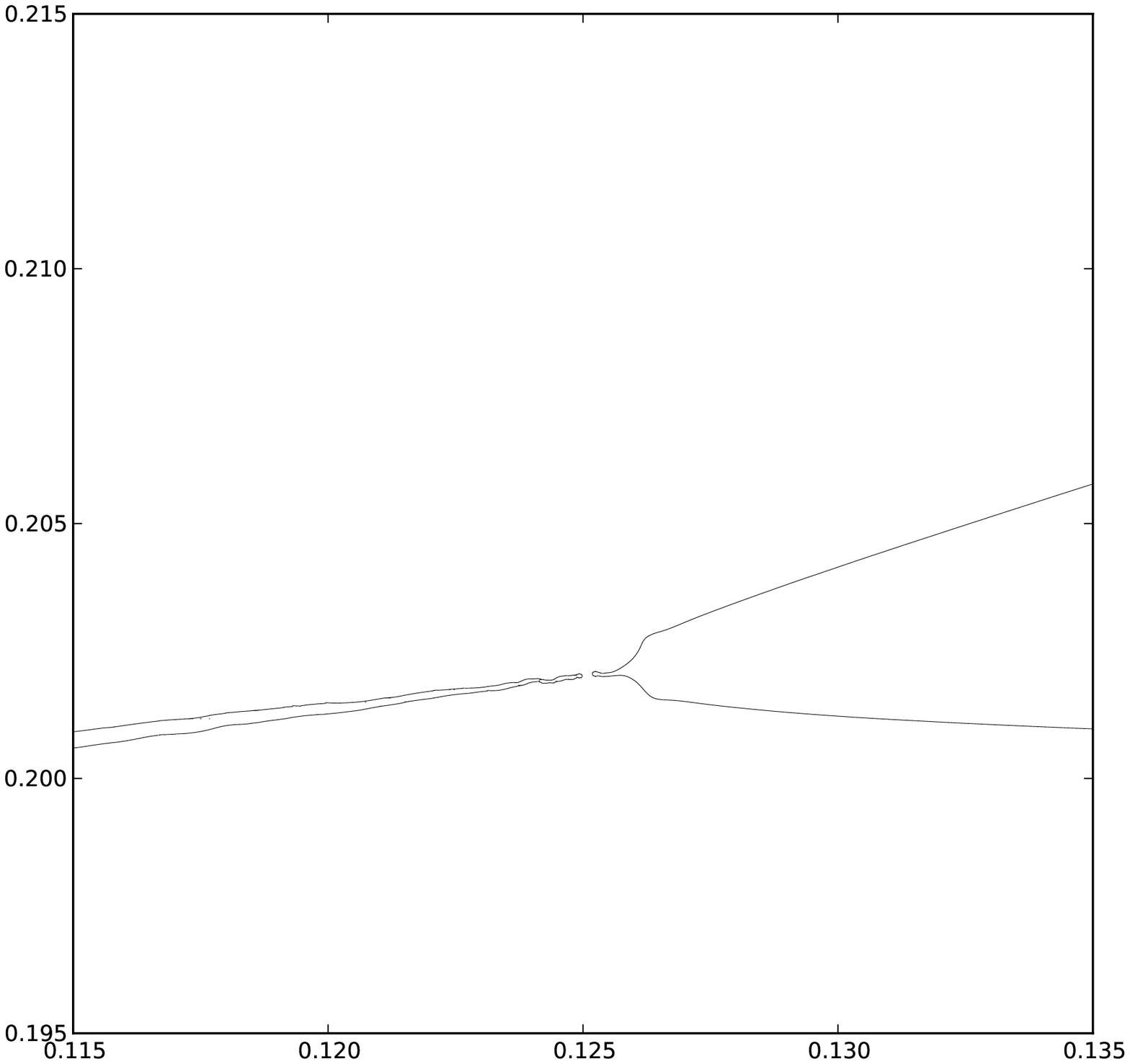}\\
c)\includegraphics[width=6. cm]{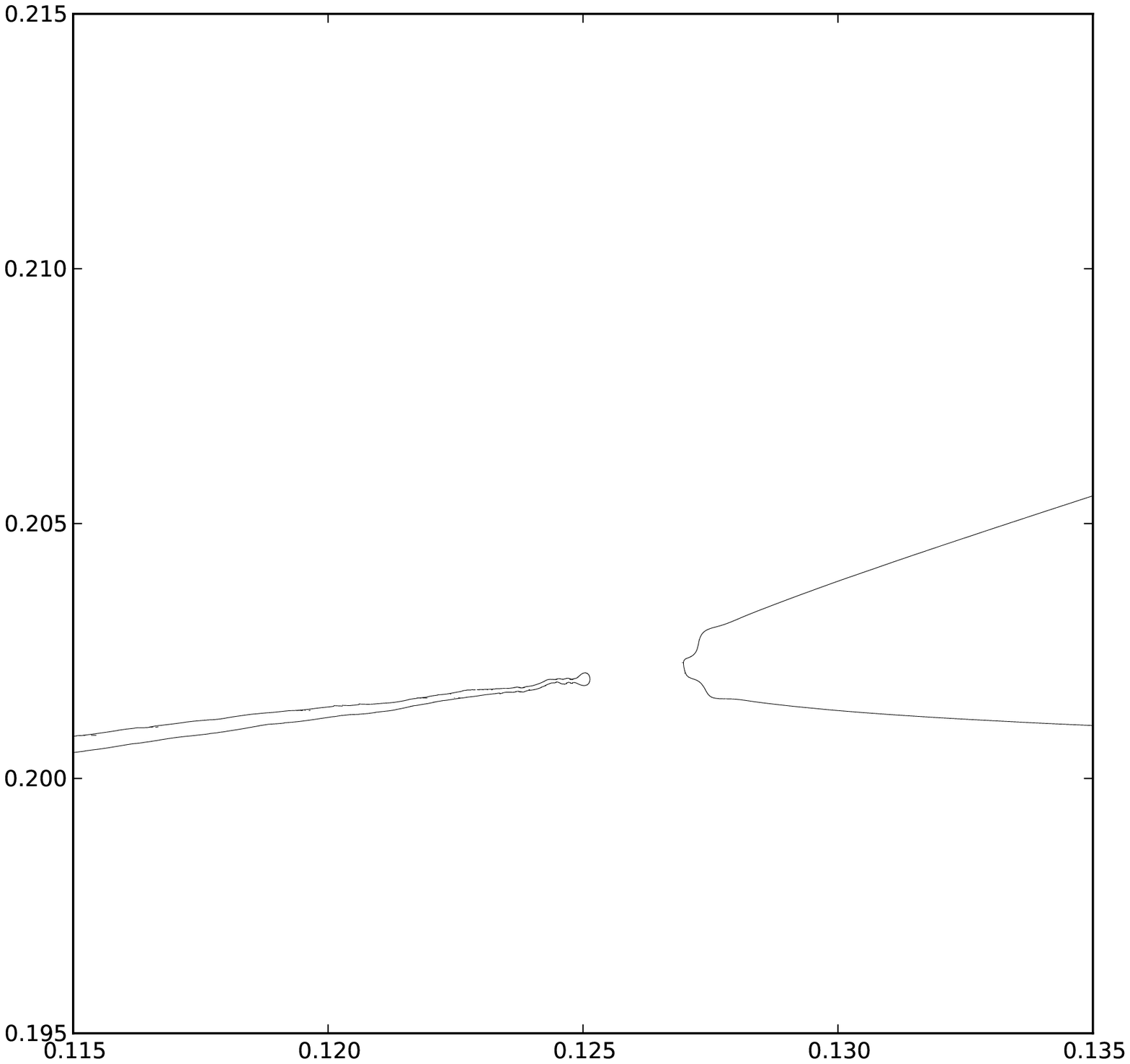}d)\includegraphics[width=6. cm]{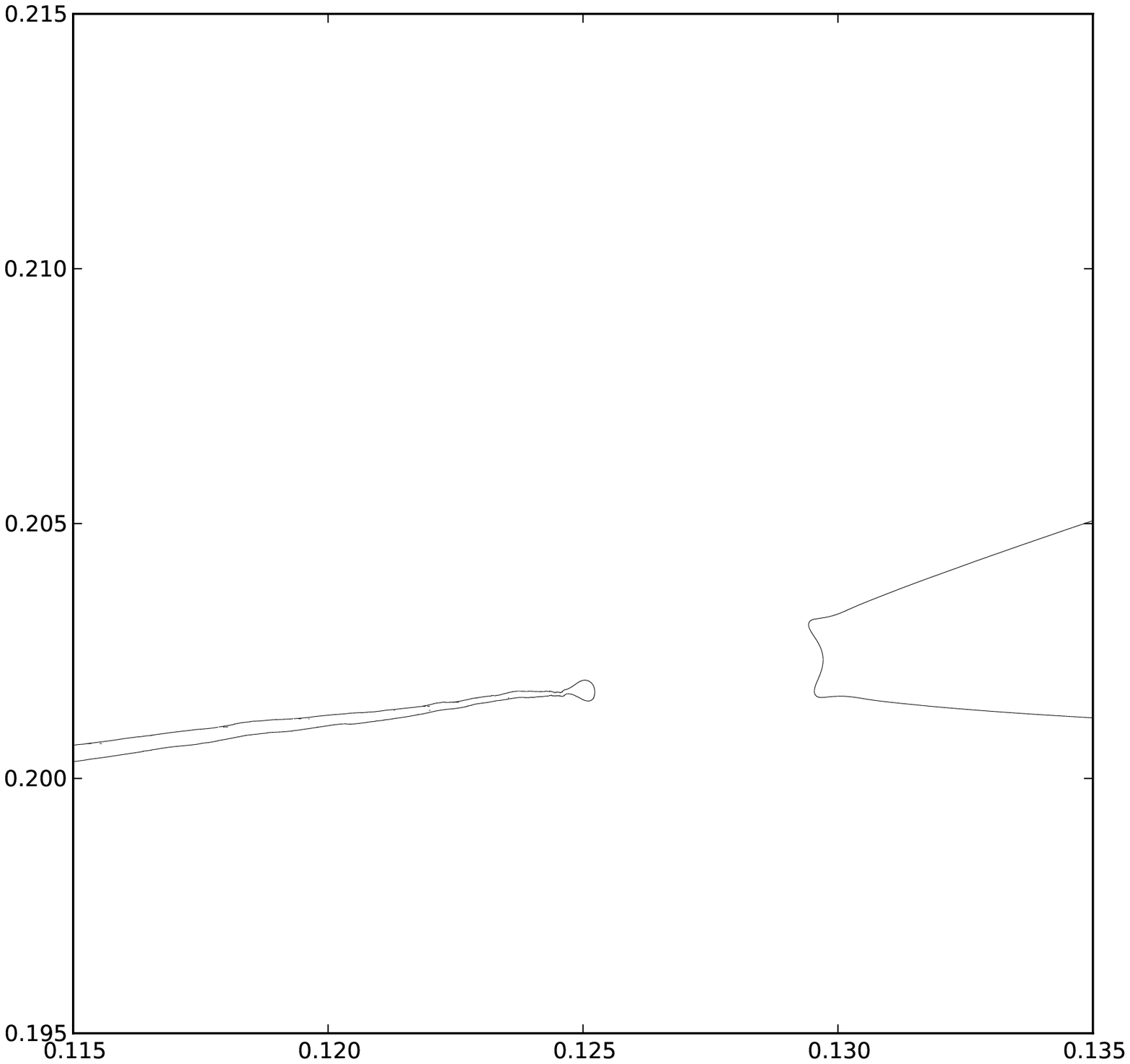}\\
\caption{Zoom on the connexion of the upper interface of the drop with the liquid layer for $\re=16000$ and $\St=2.26 \cdot 10^{-5}$, the other parameters being that of table \ref{allparams}. The interface is shown on time $U_0t/D=0.01167$ a), $0.01182$ b), $0.012$ c) and $0.0123$ d). The jet number $J=2700 \gg 1$ and the first connexion of the two interface occurs between figures b) and c). No additional bubbles are entrapped and the jet is directly emerging from this connexion, figure d).}
\label{figjet}
\end{figure}

 On the other hand, in the regime where 
  $N$ is large (and similarly $t_b \gg t_g$ it is not clear how jet formation and air bubble closing
interact but one expects that no small bubbles are entrapped and that the jet is formed simultaneously to the bubble entrapment. This is illustrated on figure~\ref{figjet} for $\re=16000$ and $\St=2.26 \cdot 10^{-7}$, so that $J=2700$.

The splashing dynamics appears thus as a complex combination between the inertial dynamics of the liquid with the cushioning of the gas.


\begin{thebibliography}{42}
\expandafter\ifx\csname natexlab\endcsname\relax\def\natexlab#1{#1}\fi

\bibitem[Culick(1960)]{Culick60}
{\sc Culick, F. E.~C.} 1960 Comments on a ruptured soap film. {\em J. Appl.
  Phys.\/} {\bf 31}, 1128--1129.

\bibitem[Deegan {\em et~al.\/}(2008)Deegan, Brunet \& Eggers]{DeBr08}
{\sc Deegan, R., Brunet, P. \& Eggers, J.} 2008 Complexities of splashing. {\em
  Nonlinearity\/} {\bf 21}, C1.

\bibitem[Duchemin \& Josserand(2011)]{DJ11}
{\sc Duchemin, L. \& Josserand, C.} 2011 Curvature singularity and film-skating
  during drop impact. {\em Phys. Fluids\/} {\bf 23}, 091701.

\bibitem[Fuster {\em et~al.\/}(2009)Fuster, Agbaglah, Josserand, Popinet \&
  Zaleski]{Fuster}
{\sc Fuster, D., Agbaglah, G., Josserand, C., Popinet, S. \& Zaleski, S.} 2009
  Numerical simulation of droplets, bubbles and waves: state of the art. {\em
  Fluid Dyn. Res.\/} {\bf 41}, 065001.

\bibitem[Gueyffier \& Zaleski(1998)]{Gue98}
{\sc Gueyffier, D. \& Zaleski, S.} 1998 Formation de digitations lors de
  l'impact d'une goutte sur un film liquide. {\em C. R. Acad. Sci. IIb\/} {\bf
  326}, 839--844.

\bibitem[Howison {\em et~al.\/}(2005)Howison, Ockendon, Oliver, Purvis \&
  Smith]{HoOc05}
{\sc Howison, S., Ockendon, J., Oliver, J., Purvis, R. \& Smith, F.} 2005
  Droplet impact on a thin fluid layer. {\em J. Fluid. Mech.\/} {\bf 542},
  1--23.

\bibitem[Jian {\em et~al.\/}(2015)Jian, Josserand, Ray, Duchemin, Popinet \&
  Zaleski]{taiwan15}
{\sc Jian, Z., Josserand, C., Ray, P., Duchemin, L., Popinet, S. \& Zaleski,
  S.} 2015 Modelling the thickness of the air layer in droplet impact. {\em
  ICLASS 2015\/} .

\bibitem[Josserand \& Thoroddsen(2016)]{JTAR16}
{\sc Josserand, C. \& Thoroddsen, S.} 2016 Drop impact on a solid surface. {\em
  Annu. Rev. Fluid Mech.\/} {\bf 48}, 365--391.

\bibitem[Josserand \& Zaleski(2003)]{JZ03}
{\sc Josserand, C. \& Zaleski, S.} 2003 Droplet splashing on a thin liquid
  film. {\em Phys. Fluids\/} {\bf 15}, 1650.

\bibitem[Klaseboer {\em et~al.\/}(2014)Klaseboer, Manica \&
  Chan]{klaseboer2014universal}
{\sc Klaseboer, E., Manica, R. \& Chan, D.~Y.} 2014 Universal behavior of the
  initial stage of drop impact. {\em Physical review letters\/} {\bf 113}~(19),
  194501.

\bibitem[Kolinski {\em et~al.\/}(2012)Kolinski, Rubinstein, Mandre, Brenner,
  Weitz \& Mahadevan]{kol12}
{\sc Kolinski, J.~M., Rubinstein, S.~M., Mandre, S., Brenner, M.~P., Weitz,
  D.~A. \& Mahadevan, L.} 2012 Skating on a film of air: drops impacting on a
  surface. {\em Phys. Rev. Lett.\/} {\bf 108}, 074503.

\bibitem[Korobkin {\em et~al.\/}(2008)Korobkin, Ellis \& Smith]{Korob08}
{\sc Korobkin, A., Ellis, A. \& Smith, F.} 2008 Trapping of air in impact
  between a body and shallow water. {\em J. Fluid Mech.\/} {\bf 611}, 365--394.

\bibitem[Lagr{\'e}e {\em et~al.\/}(2011)Lagr{\'e}e, Staron \&
  Popinet]{Lagree:2011bq}
{\sc Lagr{\'e}e, P.~Y., Staron, L. \& Popinet, S.} 2011 {The granular column
  collapse as a continuum: validity of a two-dimensional Navier--Stokes model
  with a $\mu$(I)-rheology}. {\em Journal of Fluid Mechanics\/} {\bf 686},
  378--408.

\bibitem[Lesser \& Field(1983)]{LeFi83}
{\sc Lesser, M. \& Field, J.} 1983 The impact of compressible liquids. {\em
  Annu. Rev. Fluid Mech.\/} {\bf 15}, 97.

\bibitem[Luchini \& Charru(2010)]{Luchini2010}
{\sc Luchini, P. \& Charru, F.} 2010 {Consistent section-averaged equations of
  quasi-one-dimensional laminar flow}. {\em J. Fluid. Mech.\/} {\bf 565},
  337--341.

\bibitem[Mandre \& Brenner(2012)]{Mandre12}
{\sc Mandre, S. \& Brenner, M.} 2012 The mechanism of a splash on a dry solid
  surface. {\em J. Fluid Mech.\/} {\bf 690}, 148--172.

\bibitem[Mandre {\em et~al.\/}(2009)Mandre, Mani \& Brenner]{Mandre09}
{\sc Mandre, S., Mani, M. \& Brenner, M.} 2009 Precursors to splashing of
  liquid droplets on a solid surface. {\em Phys. Rev. Lett.\/} {\bf 102},
  134502.

\bibitem[Mani {\em et~al.\/}(2010)Mani, Mandre \& Brenner]{Mani10}
{\sc Mani, M., Mandre, S. \& Brenner, M.} 2010 Events before droplet splashing
  on a solid surface. {\em J. Fluid Mech.\/} {\bf 647}, 163--185.

\bibitem[Mehdi-Nejad {\em et~al.\/}(2003)Mehdi-Nejad, Mostaghimi \&
  Chandra]{chandra03}
{\sc Mehdi-Nejad, V., Mostaghimi, J. \& Chandra, S.} 2003 Air bubble entrapment
  under an impacting droplet. {\em Phys. Fluids\/} {\bf 15}~(1), 173--183.

\bibitem[Mundo {\em et~al.\/}(1995)Mundo, Sommerfeld \& Tropea]{MST95}
{\sc Mundo, C., Sommerfeld, M. \& Tropea, C.} 1995 Droplet-wall collisions:
  Experimental studies of the deformation and breakup process. {\em Int. J.
  Multiphase Flow\/} {\bf 21}, 151.

\bibitem[Philippi {\em et~al.\/}(2015)Philippi, Lagr{\'e}e \&
  Antkowiak]{philippi15}
{\sc Philippi, J., Lagr{\'e}e, P.-Y. \& Antkowiak, A.} 2015 {Drop impact on
  solid surface: short time self-similarity}. In preparation.

\bibitem[Popinet(2003)]{Pop03}
{\sc Popinet, S.} 2003 Gerris: a tree-based adaptive solver for the
  incompressible euler equations in complex geometries. {\em J. Comp. Phys.\/}
  {\bf 190}~(2), 572--600.

\bibitem[Popinet(2009)]{GerrisVOF}
{\sc Popinet, S.} 2009 An accurate adaptive solver for surface-tension-driven
  interfacial flows. {\em J. Comput. Phys.\/} {\bf 228}, 5838--5866.

\bibitem[Popinet(2014)]{Gerris}
{\sc Popinet, S.} 2014 Gerris flow solver-http://gfs.sourceforge.net/ .

\bibitem[Rein(1993)]{Rein93}
{\sc Rein, M.} 1993 Phenomena of liquid drop impact on solid and liquid
  surfaces. {\em Fluid Dyn. Res.\/} {\bf 12}, 61.

\bibitem[Riboux \& Gordillo(2014)]{Gordillo14}
{\sc Riboux, G. \& Gordillo, J.} 2014 Experiments of drops impacting a smooth
  solid surface: A model of the critical impact speed for drop splashing. {\em
  Phys. Rev. Lett.\/} {\bf 113}, 024507.

\bibitem[Rieber \& Frohn(1998)]{Rieber98}
{\sc Rieber, M. \& Frohn, A.} 1998 Numerical simulation of splashing drops.
  Academic Press, proceedings of {ILASS98}, {M}anchester.

\bibitem[Rioboo {\em et~al.\/}(2001)Rioboo, Marengo \& Tropea]{Rio01}
{\sc Rioboo, R., Marengo, M. \& Tropea, C.} 2001 Outcomes from a drop impact on
  solid surfaces. {\em Atomization and Sprays\/} {\bf 11}, 155--165.

\bibitem[Stow \& Hadfield(1981)]{StHa81}
{\sc Stow, C. \& Hadfield, M.} 1981 An experimental investigation of fluid flow
  resulting from the impact of a water drop with an unyielding dry surface.
  {\em Proc. R. Soc. London, Ser. A\/} {\bf 373}, 419.

\bibitem[Taylor(1959)]{Taylor59c}
{\sc Taylor, G.~I.} 1959 The dynamics of thin sheets of fluid {III}.
  {D}isintegration of fluid sheets. {\em Proc. Roy. Soc. London A\/} {\bf 253},
  313--321.

\bibitem[Thoraval {\em et~al.\/}(2012)Thoraval, Takehara, Etoh, Popinet, Ray,
  Josserand, Zaleski \& Thoroddsen]{Thor12}
{\sc Thoraval, M.-J., Takehara, K., Etoh, T., Popinet, S., Ray, P., Josserand,
  C., Zaleski, S. \& Thoroddsen, S.} 2012 von k\'arm\'an vortex street within
  an impacting drop. {\em Phys. Rev. Lett.\/} {\bf 108}, 264506.

\bibitem[Thoroddsen(2002)]{tho02}
{\sc Thoroddsen, S.} 2002 The ejecta sheet generated by the impact of a drop.
  {\em J. Fluid Mech.\/} {\bf 451}, 373.

\bibitem[Thoroddsen {\em et~al.\/}(2012)Thoroddsen, Thoraval, Takehara \&
  Etoh]{thor12bubble}
{\sc Thoroddsen, S., Thoraval, M.-J., Takehara, K. \& Etoh, T.} 2012
  Micro-bubble morphologies following drop impacts onto a pool surface. {\em J.
  Fluid Mech.\/} {\bf 708}, 469--479.

\bibitem[Thoroddsen {\em et~al.\/}(2003)Thoroddsen, Etoh \&
  Takehara]{thoroddsen03}
{\sc Thoroddsen, S.~T., Etoh, T.~G. \& Takehara, K.} 2003 Air entrapment under
  an impacting drop. {\em Journal of Fluid Mechanics\/} {\bf 478}, 125--134.

\bibitem[Thoroddsen {\em et~al.\/}(2005)Thoroddsen, Etoh, Takehara, Ootsuka \&
  Hatsuki]{thoroddsen05}
{\sc Thoroddsen, S.~T., Etoh, T.~G., Takehara, K., Ootsuka, N. \& Hatsuki, A.}
  2005 The air bubble entrapped under a drop impacting on a solid surface. {\em
  Journal of Fluid Mechanics\/} {\bf 545}, 203--212.

\bibitem[Tran {\em et~al.\/}(2013)Tran, de~Maleprade, Sun \& Lohse]{Tran2013}
{\sc Tran, T., de~Maleprade, H., Sun, C. \& Lohse, D.} 2013 Air entrainment
  during impact of droplets on liquid surfaces. {\em Journal of Fluid
  Mechanics\/} {\bf 726}.

\bibitem[Tryggvason {\em et~al.\/}(2011)Tryggvason, Scardovelli \&
  Zaleski]{Tryggvason11}
{\sc Tryggvason, G., Scardovelli, R. \& Zaleski, S.} 2011 {\em Direct Numerical
  Simulations of Gas-Liquid Multiphase Flows\/}. Cambridge University Press.

\bibitem[Wagner(1932)]{Wagner32}
{\sc Wagner, H.} 1932 \"uber stoss und gleitvorg\"ange und der oberfl\"ashe von
  fl\"ussigkeiten. {\em Zeit. Angewandte Math. Mech.\/} {\bf 12}~(4), 193--215.

\bibitem[Wang {\em et~al.\/}(2012)Wang, Kurniawan \& Tsai]{Wang}
{\sc Wang, A.-B., Kurniawan, T. \& Tsai, P.-H.} 2012 About oscillation effect
  on the penetration depth of drop induced vortex ring. In {\em XXIII ICTAM,
  19-24 August 2012, Beijing, China\/}.

\bibitem[Xu {\em et~al.\/}(2005)Xu, Zhang \& Nagel]{Xu05}
{\sc Xu, L., Zhang, W. \& Nagel, S.} 2005 Drop splashing on a dry smooth
  surface. {\em Phys. Rev. Lett.\/} {\bf 94}, 184505.

\bibitem[Yarin \& Weiss(1995)]{YW95}
{\sc Yarin, A. \& Weiss, D.} 1995 Impact of drops on solid surfaces:
  self-similar capillary waves, and splashing as a new type of kinematic
  discontinuity. {\em J. Fluid Mech.\/} {\bf 283}, 141--173.

\bibitem[Yarin(2006)]{Yarin06}
{\sc Yarin, A.~L.} 2006 Drop impact dynamics: Splashing, spreading, receding,
  bouncing... {\em Annu. Rev. Fluid Mech.\/} {\bf 38}, 159.

\end{thebibliography}
\end{document}